\documentclass[sigconf]{acmart}
\AtBeginDocument{%
  \providecommand\BibTeX{{%
    \normalfont B\kern-0.5em{\scshape i\kern-0.25em b}\kern-0.8em\TeX}}}

\copyrightyear{2026}
\acmYear{2026}
\setcopyright{cc}
\setcctype{by}
\acmConference[CHI '26]{Proceedings of the 2026 CHI Conference on Human Factors in Computing Systems}{April 13--17, 2026}{Barcelona, Spain}
\acmBooktitle{Proceedings of the 2026 CHI Conference on Human Factors in Computing Systems (CHI '26), April 13--17, 2026, Barcelona, Spain}
\acmPrice{}
\acmDOI{10.1145/3772318.3791400}
\acmISBN{979-8-4007-2278-3/2026/04}
\usepackage{xspace}
\usepackage{tabularx}
\usepackage{longtable}
\usepackage{enumitem}
\usepackage{booktabs}
\usepackage{array}
\usepackage{url}
\usepackage[normalem]{ulem}
\newcommand{\Code}[1]{\textit{#1}\xspace}

\newif\ifREVISIONS
\REVISIONSfalse

\newcommand{\rev}[1]{%
    \ifREVISIONS
        \textcolor{blue}{#1}% 
    \else
        #1 
    \fi
}

\newcommand{\revtwo}[1]{%
    \ifREVISIONS
        \textcolor{red}{#1}% 
    \else
        #1 
    \fi
}

\newcommand{\revrm}[1]{%
    \ifREVISIONS
        \textcolor{red}{\sout{#1}}%
    \else
        {}%
    \fi
}

\begin{document}

\title[What We Talk About When We Talk About Frameworks in HCI]{What We Talk About When We Talk About Frameworks in HCI}

\author{Shitao Fang}
\orcid{0000-0003-1401-8482} 
\affiliation{%
  \institution{The University of Tokyo}
  %\streetaddress{Hongo 7-3-1}
  \city{Tokyo}
  %\state{Tokyo}
  \country{Japan}
  }
\email{fst@iis-lab.org}

\author{Koji Yatani}
\orcid{0000-0003-4192-0420} 
\affiliation{%
  \institution{The University of Tokyo}
  %\streetaddress{Hongo 7-3-1}
  \city{Tokyo}
  %\state{Tokyo}
  \country{Japan}
  }
\email{koji@iis-lab.org}

\author{Kasper Hornbæk}
\orcid{0000-0001-8742-1198} 
\affiliation{%
  \institution{University of Copenhagen}
  %\streetaddress{Hongo 7-3-1}
  \city{Copenhagen}
  %\state{Tokyo}
  \country{Denmark}
  }
\email{kash@di.ku.dk}


\begin{abstract}
% In HCI, frameworks serve as a type of theoretical contribution.
% They are often used in ideation, design, and evaluation. However, little is known about what constitutes a good framework.
% We surveyed 615 papers 
% From these papers, we created a taxonomy to categorize the types of framework, their roles in research, and their essential structural elements.
% Our results show that enthusiasm for proposing new frameworks exceeds the willingness to use existing ones. They also reveal a weak connection between frameworks and theory, as well as an absence of critical validation and iterative development of existing frameworks.
% Based on these findings, we propose guidelines for developing and using frameworks.

In HCI, frameworks function as a type of theoretical contribution, often supporting ideation, design, and evaluation. 
Yet, little is known about how they are actually used, what functions they serve, and which scholarly practices that shape them.
To address this gap, we conducted a systematic review of 615 papers from a decade of CHI proceedings (2015--2024) that prominently featured the term framework.
We classified these papers into six engagement types. 
We then examined the role, form, and essential components of newly proposed frameworks through a functional typology, analyzing how they are constructed, validated, and articulated for reuse.
Our results show that enthusiasm for proposing new frameworks exceeds the willingness to iterate on existing ones.
They also highlight the ambiguity in the function of frameworks and the scarcity of systematic validation. 
Based on these insights, we call for more rigorous, reflective, and cumulative practices in the development and use of frameworks in HCI.

\end{abstract}








\maketitle

\section{Introduction}

% Theoretical contributions to HCI include frameworks
Research in human-computer interaction (HCI) has always incorporated theory. 
The belief appears to be that theory helps us generalize findings from empirical studies, predict behavior with interactive systems, and understand the factors that influence such behavior. 
\citet{wobbrock16contribution}
suggested that contributions to theory consist of ``new or improved concepts, definitions, models, principles, or frameworks'' (p. 41). 
In this paper, we are particularly interested in the last form of theory, \textit{frameworks}. 

\citet{roger12HCItheory} suggested that a framework in HCI is ``a set of interrelated concepts and/or a set of specific questions that is intended to inform a particular domain area, e.g., collaborative learning, online communities or an analytic method, e.g., ethnographic studies'' (p. 4). 
In this way, frameworks are typically smaller in scope and claims than full theories, making them easier to apply. 
Frameworks are also often particularly geared towards design, informing researchers about the major dimensions in design choices, articulating important design tradeoffs, or prescribing steps in a design process. 
Frameworks may also help simply by providing a shared vocabulary, easing communication and interdisciplinary collaboration. 
Perhaps for these reasons, the literature abounds in frameworks, as seen in catalogs of them \cite[e.g.,][]{CarrollModelsTheoriesFrameworks2003} as well as in individual frameworks with a significant impact on HCI \cite[e.g.,][]{gutwinDescriptiveFrameworkWorkspace2002, mekler19meaning, 
Schneider:2018, 
Jacob08reality, blackwell2003notational}.

% Specific frameworks has been anayzed and valuable lessons learned. RBI. Trajectories. 
Yet, we know little about frameworks in HCI beyond that there are many of them and they are supposed to be important. 
For instance, the frameworks just quoted differ in whether they concern interaction, user interfaces, or the goal of HCI.
They also differ in whether they articulate stages or dimensions, describe or prescribe, or help design or analyze.  
For some frameworks, subsequent analysis has explored how they are used \cite[e.g.,][]{veltSurveyTrajectoriesConceptual2017, girouardRealityRealitybasedInteraction2019}, suggesting both easy-to-use aspects of frameworks and aspects that find little use in practice. 
Whether this characterize framework use more generally in HCI remains unclear. 
The frameworks described above also differ in how they were proposed and validated, suggesting that less is known about these aspects of frameworks than about other HCI contributions. 
 
% But we do not know, more generally, how frameworks work, whether they help the discipline, how they are used, and so on.

This paper presents a large-scale systematic review of framework-related scholarship in HCI. 
We analyze a corpus of 615 papers from a decade of CHI proceedings (2015-2024) to empirically map the landscape of framework engagement, 
deconstruct the substance of these frameworks, 
and explore the methodological practices behind them. 
We explore how frameworks are constructed and validated, and how they are articulated to be useful for the designers and researchers who seek to apply them. 
The aim of this analysis is to give more general insights into what a framework is in HCI and, from those insights, discuss how we may make frameworks more useful. 
In sum, we contributed the following.

% The present paper explores the frameworks that are used in HCI. 
% We take a sample (N=615) from a decade of papers in which the term framework figures prominently. 
% We then analyze those papers, exploring the variety of contributions that the frameworks cover and how they work with proposing or using frameworks. 
% We also explore how frameworks are constructed and validated, and how it is ensured that the frameworks that research suggest are useful to the designers or researchers that seek to use them.  
% The aim of this analysis is to give more general insights in what a framework is in HCI and, from those insights, discuss how we may make frameworks more useful. 
% In sum, we contributed the following.
\begin{itemize}
    \item A systematic mapping of how the term framework is used at CHI, identifying six primary modes of engagement and tracing their prevalence over a decade.
    \item An in-depth examination of frameworks’ defining characteristics --- functions, components, and forms --- together with an analysis of community practices on constructing, validating, and articulating frameworks for reuse.
    \item A reflection and discussion on community norms regarding the framework lifecycle, creation, and evaluation, and their relation to design.  
\end{itemize}
\section{Related Work}
First, we discuss frameworks in general; then we review work on frameworks in HCI. 
Finally, we discuss metascience on frameworks in HCI.

\subsection{Framework in Research}
Frameworks are often seen as an important part of research. Their dictionary meaning amounts to a basic structure, plan, or system of concepts, values, processes, or rules. This suggests that they often function as a guide or structure. 
% check reference more carefully 

The literature distinguishes types of framework with more specialized functions than those just discussed. They include \textit{theoretical frameworks} and \textit{conceptual framework}. 
Although the distinction is subtle, 
a theoretical framework is built upon one or more established theories, using their postulates to derive hypotheses and guide an investigation. 
A conceptual framework is a structure developed by the researcher that synthesizes concepts and relationships from various sources to contextualize the phenomenon being studied~\cite{imenda2014there, ravitch2016reason}. In the methodology literature, particular emphasis is placed on theoretical and conceptual frameworks due to their central role in structuring and directing research~\cite{luft2022literature, varpio2020distinctions}.

Another type is  \textit{methodological framework} \cite{mcmeekin2020methodological}. 
It structures the application of methods and tools in a step-by-step manner. A framework may also be a \textit{typology} or\textit{ taxonomy} \cite{bailey1994typologies}; such frameworks offer concepts to distinguish and classify phenomena, similarly organizing or structuring the world. 

Within HCI more narrowly, frameworks are also defined as an important source of inspiration and knowledge. 
\citet{wobbrock16contribution} suggested that frameworks are one type of theoretical contribution, informing ``what we do, why we do it, and what we expect from it''. 
Rogers~\cite{roger12HCItheory} defined a framework as ``a set of interrelated concepts and/or a set of specific questions that is intended to inform a particular domain area''. 

Researchers have long drawn upon, adapted, and created frameworks to organize our knowledge. 
For instance, 
Carroll et al.'s task-artifact framework~\cite{Carrol92task} extends earlier HCI traditions of task analysis and design methodology with psychology and design rationale to understand and guide the co-evolution of human tasks and technological artifacts.
Kuutti~\cite{kuutti1996activity} discussed Activity Theory from cultural–historical psychology as a framework for HCI, highlighting how interaction is a goal-directed activity mediated by tools within social and organizational contexts. 
Dey et al.~\cite{dey2001conceptual} proposed a conceptual framework for context-aware computing. The framework separates context acquisition and representation from delivery and reaction. 

These examples demonstrate quite diverse frameworks. But they also indicate that frameworks in HCI are increasingly designed to serve a dual function as both analytical lenses for understanding and generative tools for design.
For example, \citet{Benford08trajectories}'s trajectories framework helps analyze user experiences over time, while also serving as a design tool for challenges like synchronization in interactive narratives. 
Likewise, Jacob et al.'s Reality-Based Interaction (RBI) framework ~\cite{Jacob08reality} is both descriptive --- unifying interaction styles under four common themes --- and generative, guiding future interaction design by highlighting key trade-offs. 

% These examples demonstrate that frameworks are usually not just for conceptual understanding, but increasingly also actionable tools. 
This dual role aligns with \citet{bederson2003craft}'s typology of five types of theory in HCI: descriptive, explanatory, predictive, prescriptive, and generative. Frameworks may simultaneously span multiple categories.
This observation, combined with a noticeable increase in the number of papers that engage with the term ``framework'', sparked our interest.
It motivated a systematic investigation to answer the question: What is the current landscape of scholarly engagement with frameworks in HCI?

\subsection{Study and Analysis of Frameworks}

Evaluating the nature and impact of theoretical contributions is a critical activity for a research field to understand its own evolution and intellectual trajectory. 
While many frameworks have been proposed in HCI, only few papers have sought to analyze how these frameworks are used by the community.

One of the most comprehensive examples is Clemmensen et al.'s analysis of 109 papers that substantively used Activity Theory~\cite{clemmensen2016making}. 
Through a qualitative meta-synthesis, they developed a taxonomy of five purposes for which the theory was used: 
(1) as an object of analysis itself; (2) as a meta-tool to inform new tools; (3) for conceptual analysis of HCI issues; (4) for empirical analysis of phenomena; and (5) as a framework for design. 
Their key conclusion was that HCI researchers act not merely as ``theory users'' but also as ``theory-makers'', actively adapting and developing Activity Theory to fit their specific needs.

Following a similar methodology, Velt et al. conducted an analysis of 60 papers that engaged with the Trajectories framework~\cite{veltSurveyTrajectoriesConceptual2017}. Their analysis showed how researchers selectively borrowed from the framework's rich vocabulary, often ``picking and choosing'' specific concepts rather than applying the entire structure. 
% They coded the papers by the purpose trajectories served 
%(situating work, analyzing experiences, designing, or building new concepts) 
% and found that one of its most common uses was ``horizontal'' --- helping to build, critique, or extend other concepts at a similar level of abstraction.
More recently, ~\citet{girouardRealityRealitybasedInteraction2019} evaluated the impact of the RBI framework through a two-pronged approach: a content-based citation analysis of papers and a survey of HCI educators. They found that a significant portion of the citations were ``generative,'' using RBI to inspire or inform new designs. 
Their work demonstrated that RBI had become an established term and provided a clear methodology to assess both the cited (research) and uncited (educational) impact of a framework.

These studies focused on in-depth analysis of a single framework. 
While this gives insights into how individual theoretical contributions are interpreted, adapted, and utilized, it naturally leave open the question of whether the patterns observed for Activity Theory, Trajectories, or RBI are unique or if they represent broader trends in how the HCI community engages with frameworks. 

A smaller body of work has sought to define and classify the functions of frameworks from a more general perspective.
Schwarz et al. offered a commentary on the distinction between frameworks (particularly as a form of literature synthesis) and review articles~\cite{Schwarz07understanding}.
Through a synthesis of existing literature and a survey of academics, they defined the purpose of a framework article as prospective and integrative. 
They identified several key functions, including integrating previous research, defining the legitimate boundaries of a research area, and providing a new focus for future work. 
A more focused example within the HCI subfield of tangible interaction is Mazalek and van den Hoven's work mapping the space of tangible interaction frameworks~\cite{Mazalek09framing}. 
They organized frameworks along two axes, including one that classified frameworks as abstract (categorizing past systems), design (conceptualizing new systems), or build (providing concrete steps for creation).

Taken together, prior work provides two complementary perspectives: bottom-up analyzes of how specific, influential frameworks are appropriated in practice, and top-down classifications of the intended functions of frameworks. 
However, a comprehensive, data-driven typology of the frameworks proposed in the wider HCI community remains absent.
%Moreover, while we know that HCI researchers act as ``theory-makers''~\cite{clemmensen2016making}, the processes behind such creation are not well understood. 
Beyond examining how frameworks are adapted and used, we must also ask how these frameworks are constructed, validated, and articulated for serving the functions that they were intended to serve?

% Frameworks on their own are not a clearly defined entity—they blend with theories and other concepts. Authors are the first to characterize their work as a framework, but at times, others interpret them as theories. For instance, the Trajectories Conceptual Framework was positioned as a framework by its authors [8], yet as a theory by others, as reported by Velt et al. [83]. For the purpose of this current analysis, a strict definition for frameworks is not necessary. We aim to analyze the impact of a theoretical piece of research that was originally labeled a framework by its authors. Hence, we focus on frameworks, though our work may be applied to other types of theoretical works.

\subsection{Meta-Research and Literature Syntheses in HCI}

Meta-research, the study of research practices, offers valuable insights into how a field can refine its methodology, improve rigor, and address its challenges. For a ``rapidly evolving field'' like HCI, such self-reflection is critical~\cite{Oppenlaender25meta}. 

Several recent works illustrate this orientation toward disciplinary self-examination. 
Chignell et al.~\cite{Chignell23humanfactor} reviewed HCI's evolution and its relation with human factors to contextualize the field's current trajectory and future challenges with AI. 
Linxen et al.~\cite{linxen21Weird} analyzed five years of CHI proceedings to quantify the prevalence of ``WEIRD'' 
% (Western, Educated, Industrialized, Rich, and Democratic) 
participant samples, revealing significant geographic and demographic biases in HCI research practices. 
This reflexive turn also extends to analyzing the community's own outputs. 
%Stefanidi et al.~\cite{Stefanidi23review} conducted a ``review of reviews'' to map the landscape of literature synthesis papers in HCI.
Rogers et al. ~\cite{Rogers24Umbrella} performed an ``umbrella review'' specifically to assess the reporting quality of systematic review practices at CHI, identifying key areas for methodological improvement to increase transparency and rigor.
\rev{Other meta-research has studied replications \cite{hornbaek2014once}, research ethics \cite{10.1145/3544548.3580848}, positionality \cite{10.1145/3706598.3713280}, citation practices \cite{10.1145/3706598.3713556}, and the use of LLMs \cite{10.1145/3706598.3713726}.}

%Meta-research seeks to evaluate not only the content of contributions, but also the epistemic practices through which the community organizes and builds knowledge. 
%This aligns perfectly with our interest in understanding both the landscape of scholarly engagement with frameworks and the practical methods through which those frameworks are generated.

In HCI, meta-research has manifested itself as literature syntheses, surveys, taxonomies, and systematic reviews, which collectively play a crucial role in consolidating fragmented research areas and guiding future directions. 
This type of work is recognized by 
Wobbrock and Kientz as a core ``survey contribution'' in HCI~\cite{wobbrock16contribution}. 
Indeed, as established in the previous section, the foundational studies that analyze the use and impact of specific frameworks \cite[e.g.,][] {clemmensen2016making, veltSurveyTrajectoriesConceptual2017, girouardRealityRealitybasedInteraction2019, Schwarz07understanding} have relied on literature reviews as their methodology.
To produce trustworthy findings, we adopt the methodology of a large-scale Systematic Review (SR). 
This meta-research approach allows us to systematically identify and code the body of literature where ``framework'' is a central concept and to transparently report our process, ensuring a rigorous analysis of framework contributions.

\section{Method}

To investigate how the term ``framework'' is used
within HCI, we conducted a systematic review of the full papers published in the \textit{Proceedings of the ACM Conference on Human Factors}, CHI, over ten years (2015--2024). 
As a highly influential publication venue, CHI provides a representative snapshot of scholarly trends and norms.

\subsection{Research Questions}
To guide our systematic review and structure our analysis, we formulated three research questions. 
These questions allow us to chart the breadth of framework-related scholarship in HCI, examine the usage of the term, understand the framework contributions, and evaluate how these contributions are built and communicated to the community.

\begin{itemize}
    \item \textbf{RQ1}: What is the landscape of the term ``framework'' used in the recent CHI literature?
\end{itemize}

This question seeks to empirically examine the overall patterns of scholarly engagement with the term framework. 
It asks how the term is used at CHI and how its use connects to frameworks as a form of theoretical contribution.
We are interested in this question because, despite the prevalence of the term, there is no clear understanding of collective practices of our community.
Answering this will allow us to better understand the scholarly engagement and norms surrounding frameworks in HCI.

%This question aims to map the territory by qualitatively coding the different ways authors engage  with the term. Our analysis for this question will identify whether papers propose new frameworks, adapt or extend existing ones, evaluate or review a body of frameworks, apply a framework as an analytical or generative tool, or simply reference one to position their work. The details of the qualitative coding process will be described in the following section ~\ref{sec:XXX}.

\begin{itemize}
    \item \textbf{RQ2}: What types of framework are proposed in CHI, and what are their defining characteristics in terms of their function, components, and form?
\end{itemize}

This question addresses a central ambiguity: 
Although ``framework'' is a common label for research contributions, its meaning varies considerably. 
This variation can hinder effective communication within and beyond the HCI community.
By analyzing frameworks that are explicitly proposed, we aim to identify what they are meant to do (function), what they are made of (components), and how they are presented (form).
Answering this question will help us (authors, reviewers, and practitioners) develop a shared vocabulary for framework contribution.

% This question addresses the core ambiguity of the term``framework'' by seeking to build a typology of these contributions. 
% By analyzing the papers that propose new frameworks, we identify and categorize their defining characteristics based on three key facets: their intended function (e.g., conceptual, analytical, methodological), their core components (e.g., principles, dimensions, process steps), and their presentational form (e.g., visual diagrams, matrices, checklists). The resulting typology aims to create a shared vocabulary for the HCI community.

\begin{itemize}
    \item \textbf{RQ3}: How are these frameworks constructed, validated, and articulated to serve the functions they were intended to serve?
\end{itemize}

This question investigates the methodological rigor and practical utility of framework contributions. 
We are interested in this because a framework's value is not inherent; it depends on how it is constructed, how its claims are validated, and how it is articulated for others to use. 
This matters deeply to the HCI community because the answers reveal the standards to which we hold our contributions. 
Understanding these standards is essential for assessing the trustworthiness of our field's knowledge base and building cumulative knowledge from prior work.

% This final question critically evaluates the methodological rigor and practical utility of framework contributions. 
% We examine the entire lifecycle of a framework's creation and presentation, from its methodological backbone to its communication. 
% This includes analyzing the methods used for construction (e.g., literature synthesis, empirical study), the evidence provided for validation (e.g., case studies, expert reviews), and the explicit efforts authors provide to make their frameworks understandable, accessible, and usable. For articulation, we specifically look for practical guidance such as step-by-step instructions, criteria for successful application, or worked examples that enable others to apply the work.
\begin{figure*}
    \centering
    \includegraphics[width=0.6\linewidth]{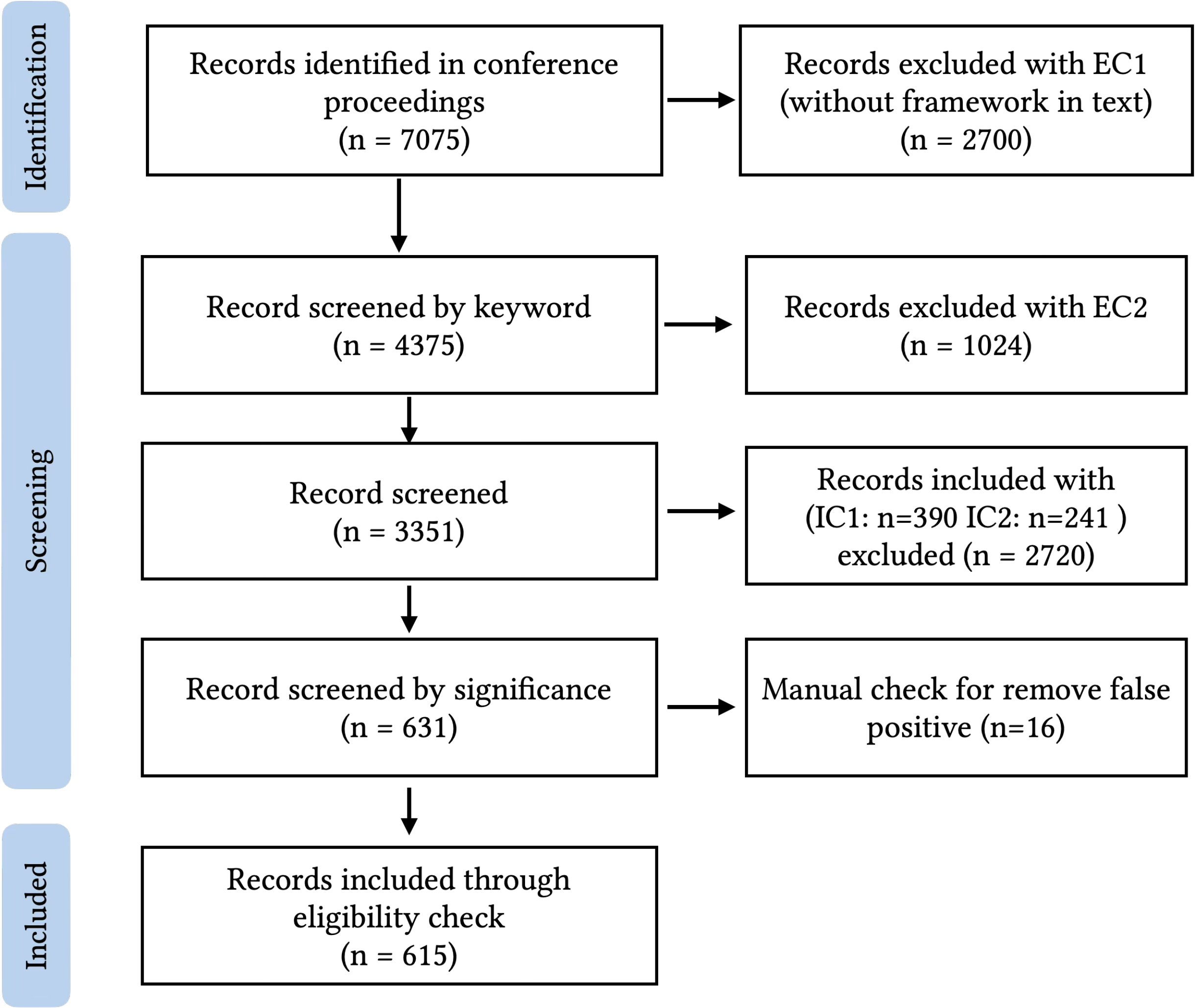}
    \caption{Adapted PRISMA flowchart demonstrating the selection process of framework-engaged paper. \rev{EC = Exclusion Criteria; IC = Inclusion Criteria.}}
    \label{fig:prisma}
    \Description{This figure is a flow diagram that outlines the process of identifying, screening, and including records in the study, similar to a PRISMA-style diagram. The process begins with 7,075 records identified from conference proceedings. At the first step, 2,700 records are excluded because they do not contain the word “framework” in the text, leaving 4,375 records to be screened by keyword. From this set, 1,024 are excluded by a second exclusion criterion, resulting in 3,351 records for further screening. Next, 2,720 of these records are excluded because they matched inclusion criteria but did not meet significance thresholds, leaving 631 records. These remaining records are screened by significance, and then a manual check is conducted to remove 16 false positives. Finally, 615 records pass the eligibility check and are included in the study. The diagram is structured vertically with boxes connected by arrows, showing the narrowing sequence from identification through screening to final inclusion. Categories such as “Identification,” “Screening,” and “Included” are labeled on the left margin to indicate the major phases of the process.}
\end{figure*}

\subsection{Paper Identification and Selection}
We followed the Preferred Reporting Items for Systematic Reviews and Meta-Analyses (PRISMA) 2020 statement~\cite{page2021prisma}. 
Our process involved a comprehensive search followed by a multi-stage, funnel-based screening process. The process is summarized in a PRISMA flow diagram (see Figure ~\ref{fig:prisma}).

\subsubsection{Sources and Data Format}
We first collected full research articles published in the proceedings of the ACM Conference on Human Factors in Computing Systems (CHI) 
\rev{for the past ten years available at the time of analysis} (2015--2024).
% \rev{We focused on CHI as a representative outlet of HCI papers, similar to other reviews \cite{Rogers24Umbrella, linxen21Weird, hornbaek2014once, 10.1145/3706598.3713280}. }
\revtwo{Similar to prior reviews~\cite{Rogers24Umbrella, linxen21Weird, hornbaek2014once, 10.1145/3706598.3713280}, we focused on CHI as a representative outlet of HCI papers.
}
% \rev{We focused on CHI as a representative outlet of HCI papers, as in prior review studies~\cite{Rogers24Umbrella, linxen21Weird, hornbaek2014once, 10.1145/3706598.3713280}.}
This initial collection, sourced from the ACM Digital Library, comprised $7,075$ papers. 
The full text of these papers was collected in two formats for analysis: For papers from 2015-2018, only PDF versions were available; for papers from 2019-2024, we collected both PDF and HTML versions, using the HTML as the primary source for its higher parsing accuracy and the PDF for cross-validation.

\subsubsection{Screening Process}
The screening process was conducted in four sequential stages to progressively refine the corpus. We applied the following exclusion (EC) and inclusion (IC) criteria:

(\textit{EC1}) - The paper does not contain the keyword ``framework'' in its full text.

(\textit{EC2}) - The paper mentions the keyword "framework" only in the reference list or appendices.

(\textit{IC1}) - The paper mentions the keyword ``framework'' at least once in the abstract or title.

(\textit{IC2}) - The paper at least mention keyword ``framework'' twice in introduction and conclusion or equivalent sections.

The screening process began with a keyword search for ``framework'' (EC1) in 7,075 CHI papers, narrowing the pool to 4,375. Papers in which the term appeared only in bibliographies or appendices were excluded (EC2), leaving 3,351 papers that engaged with the term more directly.
These papers were then screened for significance (IC1 and 2): First, by identifying 390 papers where ``framework'' was central in the title or abstract, and second, by conducting a full-text analysis to capture an additional 241 papers where the concept was crucial to the argument but not present in title or abstract.
Together, these steps produced 631 candidate papers for manual verification.

\subsubsection{Eligibility Assessing and Corpus Finalizing}

This 631 pool subsequently underwent a final manual eligibility assessment.
This step was crucial because the pipeline for processing full-text PDF was intentionally designed to be inclusive, minimizing the risk of false negatives but occasionally producing false positives. 
For example, the multi-column layouts in older PDFs could cause our parser to incorrectly attribute keywords from the session name in header or other section like ``Related Work'' to the introduction or conclusion.
Therefore, a researcher manually assessed each of the 631 papers against the inclusion criteria. This  process removed all false positives, resulting in a final validated corpus of 615 papers for analysis.

\subsection{\rev{Inductive} Coding \rev{for Engagement Type}}
To systematically answer RQ1 and inform our analysis for subsequent RQs, 
we first conducted a qualitative coding process on the corpus. 
The goal of this process was to classify each paper based on its primary mode of engagement with the term ``framework''. 
This section details the development of our codebook  and the process followed to ensure coding reliability.

\subsubsection{Codebook Development}

We developed our codebook through an iterative, grounded process. 
Two authors conducted an open-coding exercise on a random sample of 60 papers, independently identifying emergent themes that described a paper's primary engagement with the term ``framework''. 
Through regular meetings, these themes were compared, grouped, and refined until we reached theoretical saturation. 
The results were consolidated into the six primary codes: \Code{Create}, \Code{Adapt}, \Code{Validate}, \Code{Review}, \Code{Use}, and \Code{Mention} (detailed in Table~\ref{tab:codetypes}). 

\subsubsection{Coding Protocol and Principles}
\label{sec:authorcentric}

Recognizing that a single paper can exhibit multiple forms of engagement, we developed a hierarchical decision protocol to consistently assign a single primary code based on the paper's central engagement. 
The complete decision tree is available in Appendix~\ref{app:decision_tree}.

The term ``framework'' is not a clearly defined entity~\cite{girouardRealityRealitybasedInteraction2019}: 
Authors have the right to name their work, while subsequent researchers have the right to reinterpret existing knowledge as a framework for their own use. 
To overcome this and apply the codebook consistently, we established a core principle of \textit{terminological literalism} and an \textit{author-centric view}. 
Thus, coding decisions were based exclusively on the authors' language, respecting their right to both define their own contributions and reinterpret the work of others. 
The specific rules for applying this principle are detailed in Appendix~\ref{app:coding_principles}.
This approach eliminates subjectivity by removing the need to interpret an author's intent, and keep our focus on the explicit discourse surrounding the term ``framework'' itself.

\subsubsection{Coding Process and Inter-Rater Reliability}

To ensure reliability, we followed a structured coding process:

\begin{enumerate}
    \item \textbf{Rater Training}: Two primary coders were trained on the codebook definitions and decision protocol to establish a shared understanding.
    
    \item \textbf{Pilot Coding}: A random sample of 90 papers (about 15\% of the corpus) was independently coded. The coders reached 85.6\% observed agreement, with Cohen’s Kappa of $\kappa \approx 0.78$, indicating substantial agreement and showing that the initial codebook can be reliably used.
    
    \item \textbf{Refinement}: Disagreements were resolved through discussion, leading to clarifications in the codebook and protocol. A third researcher was available to adjudicate if necessary.
    
    \item \textbf{Full Corpus Coding}: One primary author then coded the entire corpus using the refined protocol, flagging any ambiguous cases for review.
    
    \item \textbf{Final Audit}: Ambiguous papers were re-reviewed, achieving 73.7\% agreement and $\kappa \approx 0.66$. This sustained substantial agreement on the most challenging cases affirmed the robustness of the protocol. All remaining disagreements were resolved by consensus.
\end{enumerate}

\subsection{\rev{Deductive Coding for Contribution Type}}

\rev{
To further understand the epistemic role of framework in knowledge production in HCI, we additionally classified the contribution type of each paper. 
We adopted the established taxonomy defined by Wobbrock and Kientz~\cite{wobbrock16contribution}, which categorizes research contributions into seven types: \textit{Empirical, Artifact, Methodological, Theoretical, Dataset, Survey,} and \textit{Opinion}.
}

\rev{
The classification was primarily based on the paper's abstract. Consistent with our engagement coding protocol, we assigned a single primary contribution type to each paper in a process similar to that used for coding engagement types. To check, two researchers independently coded a random pilot sample of 50 papers. This resulted in an inter-rater reliability of Cohen's $\kappa = 0.86$, indicating strong agreement. Following this check, the remaining papers in the corpus were coded by a single researcher.
}

\subsection{Critical Appraisal and Data Extraction}
Following the initial coding of the corpus, our analysis proceeded with an in-depth data extraction to answer our second and third research questions (RQ2 and RQ3). 
We did this through data extraction and synthesis/analysis, to be detailed next.
%This entire process was grounded in a careful reading of the full text of each paper.  
%To anchor the initial coding, we first extracted a representative quote for each paper coded Create, Adapt, Validate, Review, or Use (e.g., the specific sentence where the authors claimed to propose a framework or stated which framework guided their analysis). 
%The subsequent in-depth data extraction, detailed in the following, then focused primarily on the subset of papers coded Create or Adapt, as these represent novel framework contributions.

\subsubsection{Data Extraction Protocol}

To ensure a systematic data collection process, we developed a standardized data extraction form. 
\rev{In addition to bibliographic metadata\footnote{Including DOI, title, keywords, authors, and abstract as provided by the ACM Digital Library and exported via BibTeX.} recorded for the entire corpus,}
the level of detail extracted varied by the paper's primary code:

Full Extraction for \Code{Create} and \Code{Adapt} Papers:
For papers proposing new or adapted frameworks, we extracted a comprehensive set of properties. They were chosen to address a specific aspect of our research questions.

\begin{itemize}
\item \textbf{Framework Identity (The `What''):} To answer RQ2—What types of framework are proposed?—we first extracted data related to the framework's fundamental identity and composition. This included:
\begin{itemize}
\item \textit{Name and Purpose:} The framework's given name and the authors' exact quote defining its purpose. 
\item \textit{Claimed Type:} The category assigned by the authors (e.g., conceptual, design).
\item \textit{Core Components:} The conceptual building blocks (e.g., principles, dimensions) and visualization structure (e.g., diagram, matrix).
\item \textit{Associated Terms:} Other major terms used to describe the contribution (e.g., model, theory).
\end{itemize}

\item \textbf{Context and Scope (The ``For Whom'' and ``Where''):} To understand each framework's intended role and boundaries, we collected data on its context. This included:
\begin{itemize}
\item \textit{Intended Audience:} The target users for the framework (e.g., researchers, designers).
\item \textit{Scope and Limitations:} The stated domain of applicability and acknowledged limitations.
\end{itemize}

\item \textbf{Application Guidance (The ``How to Use It''):} To evaluate the "articulation" aspect of RQ3—How are frameworks articulated?—we collected information on the practical guidance provided to help others apply the work. This included:
\begin{itemize}
\item \textit{Application Process:} The presence and form of any method for application (e.g., steps, checklist).
\item \textit{Success Criteria:} Any criteria provided to judge a successful application.
\item \textit{Exemplars:} Specific use cases or scenarios provided for illustration.
\end{itemize}

\item \textbf{Methodology and Validation (The ``How It Was Built''):} To assess the methodological rigor of each contribution for RQ3—How are frameworks constructed and validated?—we extracted data on its development and justification. This included:
\begin{itemize}
\item \textit{Development Methodology:} The reported method for creating the framework (e.g., literature synthesis, grounded theory).
\item \textit{Evidence of Utility:} The evidence supporting the framework's value (e.g., user study, expert evaluation). 
\end{itemize}
\end{itemize}

\rev{
We performed an additional lineage extraction for \Code{Adapt}, \Code{Validate}, \Code{Review}, and \Code{Use} papers, as these all involve engagement with published frameworks. 
We recorded the specific source framework being engaged with and coded two attributes to trace their origins and adoption:
}

\begin{itemize}
    
    \item \rev{\textbf{Disciplinary Origin (Inside vs. Outside HCI):} We identified the publication venue of the original framework and classified its origin as ``Inside'' if the framework was originally published in primary HCI venues (e.g., CHI, TOCHI, CSCW, UIST, DIS, IMWUT) or books judged to be about HCI. Frameworks originating from venues outside this core list were classified as ``Outside.''}
    
    \item \rev{\textbf{Authorship Relationship (Self vs. External):} We cross-referenced the author list of each paper with that of the original framework.}
    \begin{itemize}
        \item \rev{\textit{Self-Engagement:} Coded if there was at least one overlapping author between the original framework and the current paper. This indicates a research group extending or refining their own prior work.}
        \item \rev{\textit{External Engagement:} Coded if there was no author overlap. This indicates broader community adoption, suggesting the framework possesses utility and applicability independent of its original authors.}
    \end{itemize}
    
\end{itemize}

For papers coded as \Code{Mention}, no additional data was extracted.
\rev{These instances typically represent ``signaling'' citations that acknowledge a framework for context without substantively applying it, thus offering limited value for understanding epistemic role of these frameworks.}

% Targeted Extraction for Other Categories:
% For the other categories, we extracted only the essential information needed to contextualize their role in the literature:

% \begin{itemize}
% \item For \textbf{Validate} and \textbf{Review} papers, we recorded the specific framework(s) being analyzed.
% \item For \textbf{Use} papers, we recorded the specific framework being applied.
% \item For \textbf{Mention} papers, no further data was extracted beyond the initial classification.
% \end{itemize}

\subsubsection{Synthesis and Analysis}
Inspired by the methodology of Rogers et al.'s umbrella review ~\cite{Rogers24Umbrella}, we also chose a three-stranded approach that mixes descriptive summaries, categorization, and narrative synthesis to synthesize the extracted data.

\begin{enumerate}
    \item We generated \textbf{descriptive summaries} and \rev{quantitative measures} by analyzing our extracted data across its \revrm{four} \rev{five} main dimensions (Identity, Context, Application, Methodology, \rev{and Linage}). This involved an iterative process of tagging and clustering the data to explore, for example, how authors define their framework's function and describe its core components. 
    These summaries provide a general overview of the pattern in the data. 
    \item We then performed a \textbf{systematic categorization} to add structure to these initial findings. This involved grounding our analysis in established theory \cite{bederson2003craft} to build a robust, functional typology. 
    This step allowed our data on the definition of characteristics and community practices to converge into a coherent analytical structure. 
    \rev{\item We also clustered papers in themes, so as to compared differences across areas of CHI. The details of the clustering are in \autoref{app:clustering}.}
    \item Finally, we created \textbf{narrative syntheses} from notes, observations, and discussions throughout the research process. This was essential to explore patterns that could not be easily quantified. 
\end{enumerate}

\section{Results}
% In this section, we present the results of our coding and analysis. 
% We begin by detailing the general landscape of how the term ``framework'' is used in the CHI literature to answer RQ1. 
% We then present a novel typology of the framework contributions to answer RQ2. 
% Finally, we analyze the methodological practices behind these contributions to answer RQ3.

\begin{table*}[t]
\caption{Summary of Engagement Types. Bolded words were added to highlight important terms and phrases for certain contributions.} 
\label{tab:codetypes}
    \centering
    {\scriptsize
    \begin{tabularx}{\textwidth}{p{5.2cm} X}
    \toprule
\multicolumn{1}{c}{\textbf{Type and Definition}} & \multicolumn{1}{c}{\textbf{Illustrative Quotes}} \\
\midrule

\textbf{Create} \par
Researchers \Code{Create} a framework when they explicitly structure knowledge in the form of a framework. Such papers usually claim they have proposed, developed, or contributed a framework, often referring to it as “our framework” or presenting their output as serving as a ``framework.''  &
\begin{itemize}[leftmargin=2em, labelsep=0.4em, topsep=0pt, itemsep=0pt, parsep=0pt]
\item ``\textbf{A Framework for} the Experience of Meaning in Human-Computer Interaction'' in the title of ~\cite{mekler19meaning}
\item ``We \textbf{developed a decision-making framework} explaining pregnancy loss disclosures on identified social network sites (SNS) such as Facebook'' ~\cite{10.1145/3173574.3173732}
\item ``This taxonomy can \textbf{serve as a framework} for understanding photo privacy, which can, in turn, inform new photo privacy protection mechanisms'' ~\cite{10.1145/3313831.3376498}
\item ``Specifically, this study contributes to the following areas: (1) \textbf{offering a theoretical framework} that can be used to guide the design and evaluation of learning with technologies, (2) evaluating the effects of …'' ~\cite{10.1145/3544548.3580913}
\end{itemize} \\
\addlinespace

\textbf{Adapt} \par
Researchers \Code{Adapt} a framework when they also characterize their contribution as a new framework, but explicitly claim it is derived from, extends, or refines an existing one. This involves drawing key elements from one or more established frameworks to suit a new context. &
\begin{itemize}[leftmargin=2em, labelsep=0.4em, topsep=0pt, itemsep=0pt, parsep=0pt]
\item ``We introduce the \textit{BIGexplore} framework for changing goal-oriented cases. \textit{BIGexplore} detects ... Furthermore, a user study on BIGexplore confirms that the computational cost is significantly reduced compared with the existing BIG framework, and ...''~\cite{10.1145/3491102.3517729}
\item ``The DisplayFab Roadmap: a framework to structure... DisplayFab is \textbf{derived from} identifying 4 breakpoints where the PersonalFab framework is no longer applicable to the fabrication of displays''~\cite{10.1145/3613904.3642708}
\item ``We believe our framework of adult-child complementary roles (user-observer...) contributes to the field of HCI research because it represents design partnerships in a more holistic fashion. \textbf{By augmenting} Druin’s framework on children’s roles with adults, we allow for deeper examination of ...''~\cite{10.1145/3025453.3025787}
\end{itemize} \\
\addlinespace

\textbf{Validate} \newline
Researchers \Code{Validate} a framework when it is the object of study. These papers do not produce new framework(s) but instead assess existing ones' utility, applicability, or conceptual soundness across contexts, domains, or over time. &
\begin{itemize}[leftmargin=2em, labelsep=0.4em, topsep=0pt, itemsep=0pt, parsep=0pt]
\item ``How Ready is Your Ready? \textbf{Assessing the Usability} of Incident Response Playbook Frameworks'' in the title of ~\cite{10.1145/3491102.3517559}
\item ``In this paper, we \textbf{reflect on the applicability} of the concept of trajectories to soma design. Soma design is ...''~\cite{10.1145/3313831.3376812}
\item ``As an initial exploration in bridging the theory-practice gap, we conducted a study using one well-established design framework, the Attachment Framework, \textbf{to evaluate its applicability} in use. We conducted a comparative study...'' ~\cite{10.1145/2702123.2702567}

\end{itemize} \\
\addlinespace
\addlinespace

\textbf{Review} \par
Researchers \Code{Review} a framework when they make it the object of a meta-level study. Instead of empirically testing it, they conduct a systematic survey of the literature to synthesize how a framework (or class of frameworks) has been applied, charting its influence and identifying patterns of use. &
\begin{itemize}[leftmargin=2em, labelsep=0.4em, topsep=0pt, itemsep=0pt, parsep=0pt]
\item ``A \textbf{Survey} of the Trajectories Conceptual Framework: Investigating \textbf{Theory Use} in HCI'' in the title of ~\cite{veltSurveyTrajectoriesConceptual2017}
\item ``... the design of artificial moral agents (AMAs) is an area of growing interest to HCI, and rising concern for the implications of deploying moral machines raises questions about which ethical frameworks are being used in AMAs. We performed a literature review to identify and synthesize key themes in \textbf{how} ethical theories \textbf{have been applied} to AMAs''~\cite{10.1145/3411764.3445102}
\end{itemize} \\
 \addlinespace
 \addlinespace

\textbf{Use} \newline
Researchers \Code{Use} a framework when they apply one or several existing frameworks—or other forms of knowledge they characterize as a framework—as an instrumental tool or conceptual lens to guide their research process. 
This often involves leveraging the framework(s) to structure a study, analyze data, build a system, or interpret findings. &
\begin{itemize}[leftmargin=2em, labelsep=0.4em, topsep=0pt, itemsep=0pt, parsep=0pt]
\item ``Order in the Warez Scene: Explaining an Underground Virtual Community \textbf{with} the CPR Framework'' in the title of ~\cite{10.1145/2858036.2858341}
\item ``\textbf{Using} trajectories \textbf{as an analytical framework} for the audience ‘journey’ made apparent: how the ...'' ~\cite{10.1145/3173574.3174142}
\item ``We \textbf{use the lens} of feminist HCI in the process of analyzing and discussing our findings. The feminist HCI framework as described by Bardzell suggests a commitment to feminist as well as scientific objectives'' ~\cite{10.1145/3025453.3025532}
\item ``Using the Internet-enhanced self-disclosure hypothesis \textbf{as a framework}, we conducted an online survey of 274 Grindr users''~\cite{10.1145/3025453.3025775}
\end{itemize} \\
\addlinespace
\addlinespace

\textbf{Mention} \newline
Researchers \Code{Mention} a framework when none of the activities above are involved. This includes citing existing frameworks to provide background knowledge, identify a research gap, or position the current study. It also includes using the term in a general sense, without connection to a specific work, such as calling for the creation of a future framework or referring to a general class of tools. 
&
\begin{itemize}[leftmargin=2em, labelsep=0.4em, topsep=0pt, itemsep=0pt, parsep=0pt]
% \item ``In recent years there has been consideration of how to formally integrate trauma into technology design processes, leading to frameworks and guidelines for adapting the trauma-informed approach (TIA) from public health and social work to HCI.''~\cite{10.1145/3613904.3642045}
\item ``However, despite the availability of context-detectors and programming frameworks for defining how such applications should trigger, designers lack ...''~\cite{10.1145/3491102.3501902}
\item ``Results show clear evidence ...; and urge adoption of a multi-level framework for understanding ambiguity that also includes private information and infrastructure-level attributes of interaction media''~\cite{10.1145/2702123.2702368}
\item ``The comparison between human based and automatic results also revealed a complex
framework: algorithms were better or as good as human experts at evaluating webpages on specific guidelines'' ~\cite{10.1145/3290605.3300738}
\end{itemize}\\
\addlinespace

\bottomrule
    \end{tabularx}
    }
\end{table*}

\subsection{Six Types of Engagement with Frameworks}

Based on our coding of CHI papers, we identified six types of engagement with the term ``framework.'' 
Below we define each type, report the number of instances, and provide illustrative examples.

\begin{figure*}
    \centering
    \includegraphics[width=1\linewidth]{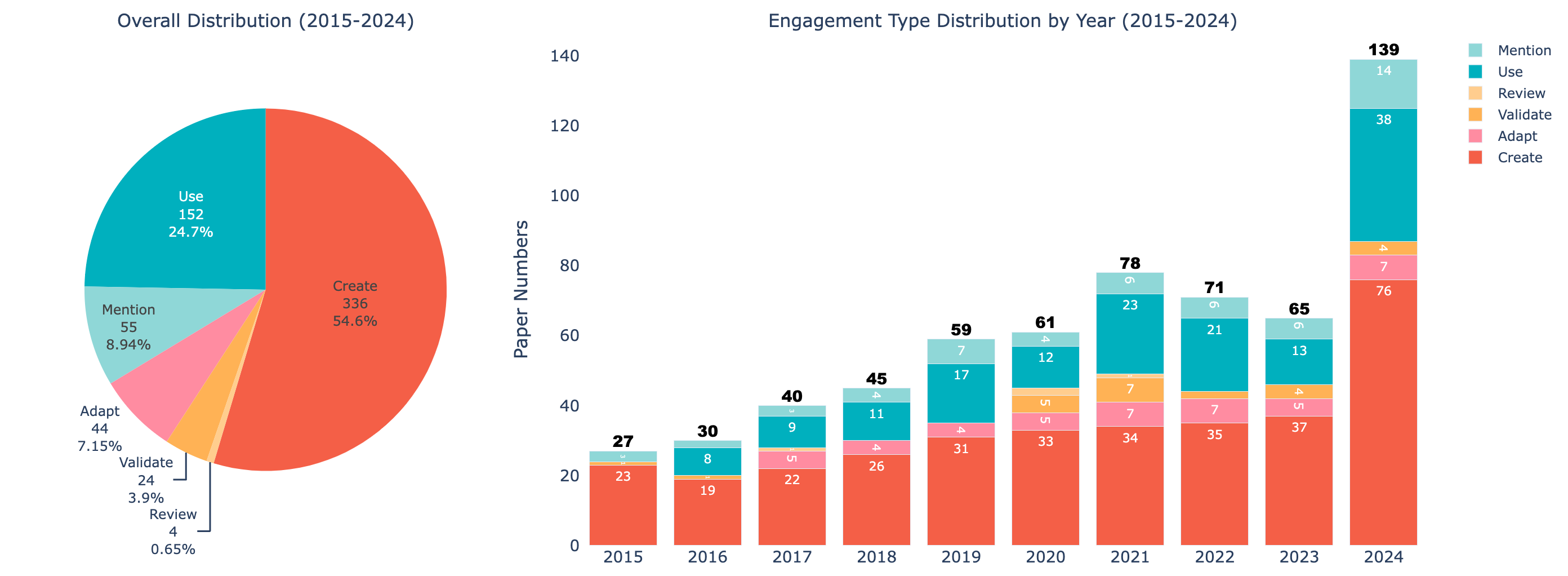}
    \caption{Six types of engagement with frameworks (N = 615). }
    \label{fig:codingresults}
    \Description{The figure presents two panels that summarize how CHI papers from 2015 to 2024 engaged with frameworks. On the left, a pie chart shows the overall distribution across 615 papers. More than half of the papers, 336 or about 55 percent, are classified as “Create,” meaning they introduce new frameworks. The next largest group is “Use,” with 152 papers or about 25 percent. Smaller proportions fall into other categories: 55 papers or 9 percent are “Mention,” 44 or 7 percent are “Adapt,” 24 or 4 percent are “Validate,” and only 4 papers, less than 1 percent, are “Review.” This distribution highlights that creating new frameworks is by far the dominant form of engagement, while validation and review are rare.
    On the right, a stacked bar chart traces these engagement types year by year. In 2015, there were 27 papers, most of them in the “Create” category. The yearly totals grow gradually, reaching 59 papers in 2019 and then climbing more steeply in the early 2020s. The peak comes in 2024 with 139 papers, more than double the number from just a few years earlier. In that year, 76 papers created new frameworks and 38 used existing ones, while the remaining categories—mention, adapt, and validate—each accounted for only a handful. Across the entire decade, the same pattern holds: most papers introduce frameworks, far fewer reuse, adapt, or evaluate them, and review papers are almost nonexistent.
    Overall, the figure conveys that framework engagement at CHI has expanded rapidly over the past ten years, but this growth has been driven overwhelmingly by the creation of new frameworks rather than by cumulative practices such as validation, adaptation, or review.}
\end{figure*}

\subsubsection{Create}

The most common form of engagement in the corpus was \Code{Create}, accounting for 336 papers, or 54.6\% of our corpus. Researchers \Code{Create} a framework when they structure knowledge into a novel conceptual artifact and present it as a primary contribution.

In practice, authors signal this contribution in two primary ways.  
The most direct is an explicit statement of creation, using verbs like ``develop'' \cite[e.g.,][]{10.1145/3173574.3173732, 10.1145/3491102.3517699, 10.1145/3613904.3641953}, ``propose'' \cite[e.g.,][]{10.1145/3025453.3025791, 10.1145/3173574.3174209, 10.1145/3290605.3300831}, ``introduce''~\cite{10.1145/2702123.2702232, 10.1145/3313831.3376826, 10.1145/3613904.3642943}, and ``present'' \cite[e.g.,][]{mekler19meaning, 10.1145/3290605.3300624, 10.1145/3313831.3376460}.
Another approach is to suggest the framework as an outcome, such as a taxonomy \cite[e.g.,][]{10.1145/3313831.3376498, 10.1145/3544548.3581126} or a design space \cite[e.g.,][]{10.1145/3290605.3300503} which can serve as a framework. 
We also noted a rhetorical move where the framework is introduced at the end of the introduction as one of the paper's key contributions. For instance, one paper wrote that 
``this study contributes to the following areas: (1) offering a theoretical framework...(2)...(3)...''~\cite{10.1145/3544548.3580913}. 
A few papers use the paper's title to directly name their framework \cite[e.g.,][]{mekler19meaning, 10.1145/2858036.2858577}.
% For example, ``A Framework for the Experience of Meaning in Human-Computer Interaction''~\cite{mekler19meaning} or ``Inspect, Embody, Invent: A Design Framework for Music Learning and Beyond''~\cite{10.1145/2858036.2858577}. 

\subsubsection{Adapt}
The second form of a framework contribution is \Code{Adapt}, appearing in 44 papers (7.15\%). 
Researchers \Code{Adapt} a framework when they characterize their contribution as a new framework but explicitly claim that it is derived from, extends, or refines an existing one. 

Note that these counts follow an author-centric view: not all adaptation activities are coded as \Code{Adapt}, 
but only those in which authors explicitly identify the contribution they build on as a ``framework.'' 
For example, papers that adapt from existing models~\cite{10.1145/3491102.3502207}, theories~\cite{10.1145/3544548.3580644}, or methods~\cite{10.1145/3544548.3580648} in order to propose a new framework are classified as \Code{Create}.

% Note that this is number after applied a terminological literalism principle, only focusing on the the term ``framework'' and author’s directly framing. 
% Some papers adapt existing models or theories are classified to be \Code{Create} (e.g. ~\cite{10.1145/3491102.3502207}). 

Similarly to \Code{Create}, these contributions are presented as novel, but identifying them requires a more careful reading of the text for keywords that signal an evolutionary process, such as ``derived from''~\cite{10.1145/3613904.3642708}, ``by augmenting''~\cite{10.1145/3025453.3025787} or ``integrating concepts'' from a framework source~\cite{10.1145/3290605.3300543}. 
We observed a practice of naming these adapted frameworks in ways that acknowledge their lineage; for example, the \textit{BIGexplore} framework~\cite{10.1145/3491102.3517729} clearly signals its relationship to the existing \textit{Bayesian information gain (BIG)} framework. 
The \textit{f-MDA} framework~\cite{10.1145/3491102.3517721} similarly modified the \textit{Mechanics-Dynamics Aesthetics (MDA)} framework~\cite{hunicke2004mda}
by adding fabrication components. 
% This engagement also create a clear line of scholarly adaptation over time. 

\begin{figure*}
    \centering
    \includegraphics[width=1\linewidth]{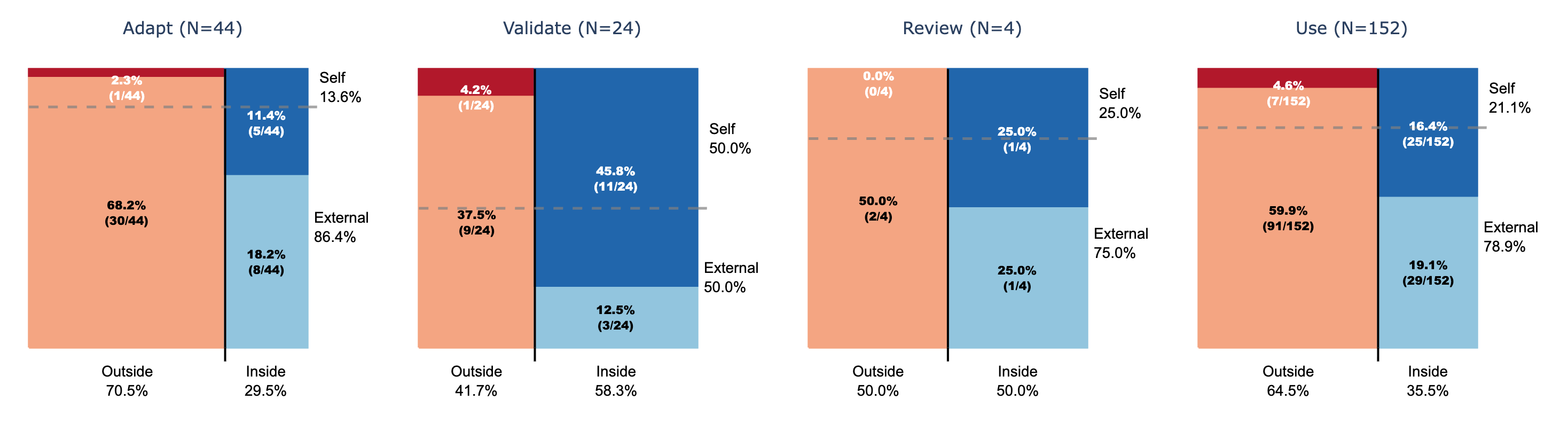}
    \caption{\rev{Mosaic plots showing the conditional probability of authorship given the disciplinary origin in the engagement types \Code{Adapt}, \Code{Validate}, \Code{Review}, and \Code{Use}. The width of each bar represents the proportion of frameworks originating Outside  vs. Inside HCI. The vertical split within each bar represents the proportion of Self vs. External engagement. The dashed line indicates the global average of External engagement for that category.}}
    \label{fig:linage}
    \Description{The figure shows four side-by-side stacked bar charts for Adapt, Validate, Review, and Use. Each chart is divided into contributions from work originating outside or inside HCI, and further split into self-authored and externally authored frameworks.  Percentages and counts indicate how often each category appears. Across all activities, most frameworks come from outside HCI and are authored by others rather than the researchers using them.}
\end{figure*}

\rev{
In our lineage analysis, as shown in Figure~\ref{fig:linage}, we trace the origins of the source frameworks being adapted. 
Our analysis shows that adaption most frequently draws on a framework from outside of HCI. 
The majority of adapted frameworks (70.5\%) originates \textit{Outside HCI} (e.g., psychology, sociology, or management science). Furthermore, 86.4\% of the papers coded as adapt represent \textit{external engagement}, which means authors are adopting and modifying frameworks created by independent researchers rather than iteratively refining their prior work.
}

\subsubsection{Validate}
A rarer form of engagement was \Code{Validate}, found in only 24 papers (3.9\%). Researchers \Code{Validate} a framework when the framework itself becomes the object of study.
They assess its properties in different fields, new contexts, or time, 
as frameworks are often shaped by specific domains, assumptions, and historical constraints. 
One typical practice for \Code{Validate} involves taking an established framework and conducting a study to ``evaluate its applicability in use'' to bridge the gaps between theory and practice \cite[e.g.,][]{10.1145/2702123.2702567}. 
%, such as the Attachment Framework~\cite{Odom09CHIattach},
\rev{
The majority of validated frameworks (58.3\%) originate from \textit{Inside HCI}. The rate of \textit{self-engagement} is notably higher than in \Code{Adapt}; specifically, 11 out of the 24 papers represent instances where HCI researchers are refining or stabilizing their own previously contributed knowledge.
}

Some work demonstrates how validation can affirm the usefulness of earlier frameworks in new contexts.
For instance, Pyle et al.~\cite{10.1145/3411764.3445331} took the Disclosure Decision-Making framework~\cite{10.1145/3173574.3173732}, originally developed to explain pregnancy loss disclosures on Facebook, and applied it to LGBTQ individuals’ disclosure practices. Their findings confirmed that the six factors in the original framework remained relevant.
Elkhuizen et al.~\cite{10.1145/3613904.3642142} validated the Materials Experience framework \cite{10.1145/2702123.2702337} by operationalizing its four experiential levels in the domain of cultural heritage. Their study of physical and virtual pop-up books confirmed that the framework provided a comprehensive structure for analyzing material engagement. 
In particular, we observed a direct lifecycle within our corpus, as both the Disclosure Decision-Making ~\cite{10.1145/3173574.3173732} and Materials Experience~\cite{10.1145/2702123.2702337} frameworks were originally proposed in papers we coded as \Code{Create}.

% For instance, such as applying the Disclosure Decision-Making framework~\cite{10.1145/3173574.3173732} to LGBTQ pregnancy loss disclosures~\cite{10.1145/3411764.3445331} or using the Materials Experience Framework~\cite{10.1145/2702123.2702337} to analyze cultural heritage artifacts~\cite{10.1145/3613904.3642142}. 

At the same time, \Code{Validate} often reveals gaps and shortcomings, as in studies showing how privacy frameworks overlook certain requirements~\cite{10.1145/3313831.3376167} or how existing canons must be expanded through afrofuturism and feminist perspectives~\cite{10.1145/3491102.3502118}. 
This highlights the key difference to \Code{Adapt}: While a validation study may conclude with suggestions for how a framework could be modified, its primary task is systematic evaluation, not the proposal of a new, fully-formed adaptation of a framework.

\subsubsection{Review}
Even more uncommon was \Code{Review}, with four instances (0.65\%), making it the rarest form of engagement.
Researchers engage in \Code{Review} when they make framework the
object of a meta-level study.
Unlike \Code{Validate}, which empirically probes whether a framework works in a specific context, \Code{Review} synthesizes how a framework (or a class of frameworks) has been used across bodies of work, mapping patterns of influence and charting their roles in research practice.

We saw two types of review. 
The first involves direct reviews of the usage of the framework, in which the framework itself is investigated. 
For example, Velt et al.~\cite{veltSurveyTrajectoriesConceptual2017} systematically analyzed 60 papers citing the trajectories framework, revealing how different parts were reused and whether its uptake aligned with its stated goals. 
Similarly, Zoshak and Dew \cite{10.1145/3411764.3445102} reviewed 53 papers on artificial moral agents to synthesize the ethical frameworks used (e.g., deontology, consequentialism, Confucianism), highlighting dominant trends and neglected alternatives.

The second form is an indirect review, where frameworks are not the initial subject but instead a key finding that emerges from a broader systematic analysis of a topic. 
% The second form involves reviews that examine frameworks behind system design or study practices. 
% Here, frameworks are not the sole object, but emerge as critical analytic layers within a broader scoping or systematic review. 
For instance, Prpa et al. ~\cite{10.1145/3313831.3376183} analyzed 31 breath-based interactive systems, identifying and reflecting on four theoretical frameworks---breathing regulation, mindfulness, soma design, and social connection---that guided the design of those systems. 
Baykal et al. \cite{10.1145/3313831.3376291} conducted a systematic review of collaborative technologies for children with special needs, documenting the evaluation criteria and theoretical frameworks that shape collaboration research.

Although rare, \Code{Review} plays a vital role in the ecology of frameworks.  By going beyond individual validations, review work provides higher-level reflexivity. It illuminates how frameworks circulate, where they structure research practice, and how their conceptual standing evolves within HCI.

\subsubsection{Use}
A total of 152 papers (24.7\%), fell into the \Code{Use} category, the second-most common form of engagement. 
Here, researchers \Code{Use} existing frameworks---or knowledge they themselves characterized as frameworks---to guide their research. 
This use occurs at all stages of the research process, such as designing studies, analyzing data, interpreting findings, and building systems.
\rev{Our lineage analysis in Figure~\ref{fig:linage} reveals a pattern similar to that of \Code{Adapt}: The source frameworks utilized in these papers mainly originate from \textit{Outside HCI} (64.5\%). Totally, 78.9\% of these cases represent \textit{external engagement}, indicating that researchers are primarily adopting tools created by the broader community rather than applying their own prior work.}

\Code{Use} reflects the dual nature of the term framework. 
The framework being used may be either knowledge explicitly claimed as a framework by its original authors or it may be a conceptual tool that the citing authors treat as a framework in their own work.
In practice, such engagement is highly diverse. 
For example, one study explicitly placed the trajectories as an ``analytical framework'' for examining audience experiences ~\cite{10.1145/3173574.3174142}, while another invoked feminist HCI as a framework to shape the interpretation of the findings~\cite{10.1145/3025453.3025532}. 
Still others described applying the Internet-enhanced self-disclosure hypothesis ``as a framework'' to design and interpret survey research on online disclosure ~\cite{10.1145/3025453.3025775}.

These practices can be seen as catering to the disciplinary expectation that good research should be grounded theoretically. This might happen in part by calling something a framework. 
At the same time, they demonstrate the instrumental role of frameworks in HCI research. 
Frameworks help structure the research design, provide analytical coherence, and connect findings to broader theoretical conversations. They are not merely background references.

It is worth noting that \Code{Use} a framework often co-occurs with other engagement types. 
For example, a paper simultaneously \Code{Created} a methodological framework to integrate user self-reports with psychophysiological data and \Code{Used} the User Action Framework to analyze their data from different evaluators~\cite{10.1145/2858036.2858511}. 
Such overlaps are common: papers that \Code{Adapt}, \Code{Validate}, or \Code{Review} frameworks often also use additional frameworks along the way. 
In our coding, we only identified the paper as \Code{Use} when no other engagement types were found. 
%Thus, the frequency m
%We believe that \Code{Use} is a more widely distributed type of engagement in the community.

% Although a wide variety of frameworks appeared under this code, we also observed a small number of frameworks that were repeatedly used in multiple papers (see Table~\ref{tab:frameworks used}). 
% \rev{Most of these frameworks are published \textit{Inside HCI} or \textit{Outside HCI} but strongly relate to the `third wave'' of HCI~\cite{threewave}, which encompasses broader social, cultural, ethical, and emotional aspects. }

Although a wide variety of frameworks appeared under this code, \rev{we observed a long tail distribution where} only a small number of frameworks were repeatedly used across multiple papers (see Table~\ref{tab:frameworks used}). 
\rev{Regardless of whether they originated \textit{Inside} or \textit{Outside} the field, these frequently used frameworks belongs with the ``third wave'' of HCI~\cite{threewave}, encompassing broader social, cultural, ethical, and emotional dimensions.}

% \revrm{These frameworks well reflect the current ``third wave'' of HCI~\cite{threewave}, which encompasses broader social, cultural, ethical, and emotional aspects.}

% While a wide variety of frameworks appeared under this code, we also noted a small number of particularly influential frameworks that were repeatedly used across multiple papers (see Table~\ref{tab:frameworks used}). 
% Surprisingly we also witnessed a small piece of framework life cycle within our corpus, one 

\begin{table*}[ht]
\centering
\begin{tabular}{p{8cm} p{7cm}}
\toprule
\textbf{Framework} & \textbf{Papers} \\
\midrule
% For Create: \cite{smith2020}\textsuperscript{*}
% For Adapt: \cite{jones2021}\textsuperscript{$\dagger$}
% For Validate: \cite{lee2022}\textsuperscript{$\ddagger$}
% For Review: \cite{kim2023}\textsuperscript{$\S$}

Contextual Integrity (CI) framework & 
\cite{10.1145/3613904.3642645}, 
\cite{10.1145/3544548.3581376}, 
\rev{
\cite{10.1145/3313831.3376167}\textsuperscript{$\ddagger$},
}
\cite{10.1145/3491102.3501883},
\rev{
\cite{10.1145/3313831.3376573}\textsuperscript{$\dagger$},
}
\cite{10.1145/3544548.3581181},
\cite{10.1145/3411764.3445517},

 \\

Feminist HCI framework& 
\cite{10.1145/2858036.2858409}, 
\cite{10.1145/3025453.3025532}, 
\cite{10.1145/3313831.3376315}\textsuperscript{*},
\cite{10.1145/3290605.3300302}

%this one is Create, I wonder if we should include it， it has no relationship with their proposed framework, but they use some structure in feminist HCI framework
\\

Value Sensitive Design (VSD) framework & 
\cite{10.1145/3613904.3642470},
\cite{10.1145/3411764.3445168}, 
\cite{10.1145/3613904.3642810}\textsuperscript{$\ddagger$}
\\

\rev{Soma Framework} &
\rev{
\cite{10.1145/3313831.3376183}\textsuperscript{$\S$},
\cite{10.1145/3411764.3445482}\textsuperscript{$\ddagger$},
\cite{10.1145/3313831.3376812}\textsuperscript{$\ddagger$}
}\
\\

Mechanics-Dynamics Aesthetics (MDA) framework & 
\cite{10.1145/3173574.3173946}, 
\cite{10.1145/3491102.3517721}\textsuperscript{$\dagger$},
\cite{10.1145/3173574.3173760}

\\

Trajectories framework & 
\cite{10.1145/3173574.3174142}, 
\cite{10.1145/3411764.3445482}\textsuperscript{$\ddagger$},
\cite{veltSurveyTrajectoriesConceptual2017}\textsuperscript{$\S$}

\\

Fleck and Fitzpatrick’s reflection framework & 
\cite{10.1145/3411764.3445670},
\cite{10.1145/3491102.3501991}\textsuperscript{$\dagger$},
\cite{10.1145/3411764.3445112}
 \\

\rev{Tronto's Care Ethics framework} &
\rev{
\cite{10.1145/3544548.3580831},
\cite{10.1145/3613904.3642614},
\cite{10.1145/3491102.3501956}
}

\\

\rev{Davis’ Mechanisms and Conditions framework} &
\rev{
\cite{10.1145/3491102.3517513},
\cite{10.1145/3613904.3642690}
}
\\

Activity Theory framework & 
\cite{10.1145/3411764.3445272}, 
\cite{10.1145/3025453.3025987} 
\\

\rev{
Teen Online
Safety Strategies (TOSS) framework} &
\rev{
\cite{10.1145/3613904.3642567},
\cite{10.1145/3173574.3174097}
}
\\

Intersectionality framework &
\cite{10.1145/3290605.3300369}, 
\cite{10.1145/3025453.3025766}\textsuperscript{*} %this one is also create
\\

Environmental Stewardship framework &
\cite{10.1145/3544548.3581493}\textsuperscript{$\dagger$},
\cite{10.1145/3491102.3517679}\textsuperscript{$\dagger$}

\\  

\bottomrule
\end{tabular}

\caption{\rev{Frequently engaged frameworks and associated papers. Citations without markers indicate a \Code{Use} engagement. Other engagement types are denoted as follows: \Code{Create} ($*$), \Code{Adapt} ($\dagger$), \Code{Validate} ($\ddagger$), and \Code{Review} ($\S$).}}

% Frequently used frameworks and associated papers. (*) indicate engagements other than \Code{Use} (e.g., Create, Adapt, Validate, Review).}
\label{tab:frameworks used}
\end{table*}

\subsubsection{Mention}

Finally, \Code{Mention} was a frequent, lightweight form of engagement, present in 55 papers (8.94\%). 
Researchers \Code{Mention} a framework when none of the more substantive codes are involved. 

In practice, \Code{Mention} includes a range of light-touch engagements (see quotes from Table~\ref{tab:codetypes}): 
Some papers cite specific frameworks to situate their work (e.g., noting the availability of programming frameworks for defining application triggers~\cite{10.1145/3491102.3501902}); others call for the creation of new frameworks (e.g., urging adoption of a multi-level framework for understanding ambiguity~\cite{10.1145/2702123.2702368}); and still others use the term in a purely rhetorical sense, such as describing a ``complex framework'' to characterize the pattern of results when comparing human-based and automatic evaluations of webpages~\cite{10.1145/3290605.3300738}.

% In practice, mentioning includes citing frameworks to provide background knowledge, calling for the creation of new frameworks, or using the term in a purely rhetorical way.

% Some papers cited specific frameworks to situate their work (e.g., noting the availability of programming frameworks for defining application triggers~\cite{10.1145/3491102.3501902}). 
% Others called for the development of new frameworks (e.g., urging adoption of a multi-level framework for understanding ambiguity~\cite{10.1145/2702123.2702368}).
% Still others used the term framework to gesture towards a general class of tools or approaches (e.g., trauma-informed design frameworks~\cite{10.1145/3613904.3642045}).

Note that this type plays a relatively minor role in shaping the engagement with frameworks  overall. 
Yet, in some citation analyses, for example, the Trajectories framework~\cite{veltSurveyTrajectoriesConceptual2017} and the Reality-based Interaction framework~\cite{girouardRealityRealitybasedInteraction2019} usage, such mentions can be important, as they are often disaggregated to show fine-grained patterns of citation.
This level of detail, however, was not our focus and as we only included papers that mentioned framework more than twice.

\subsection{The Landscape of Framework \rev{Term} Usage (RQ1)}

With the six types of engagement clarified, we can now examine how the term ``framework'' circulates throughout a decade of CHI publications. 
Specifically, we quantify its prevalence over time, analyze where it appears in papers, and summarize the aggregate composition of engagement practices. 
% The detailed ``framework'' usage position and their engagement type across year is showed in the Appdendix~\ref{app:comprehensiveanalysis}. 

% Table~\ref{tab:comprehensive_framework} shows the term ``framework'' usage position and their engagement type across year. 

\subsubsection{Rapid Growth in Recent Publications}

First, our data reveal a dramatic increase in the volume of what we term ``framework-engaged'' literature, papers that feature the term ``framework'' in their title, abstract, or key framing sections. 
As shown in Figure~\ref{fig:codingresults}, the absolute number of these articles in our corpus increased more than fivefold over the decade, from 27 papers in 2015 to 139 in 2024. 
This trend significantly outpaces the general growth of CHI; 
while the total number of publications at CHI has also risen, the proportion of framework-engaged papers has roughly doubled. 
This body of work has grown from representing approximately 5\% (27/484) of all CHI publications in the early years of our sample to over 10\% (139/1057) in 2024.
This growing prominence suggests that ``framework'' is not only an increasingly favored term among researchers but is also solidifying its role as a central and legitimate form of contribution within the HCI community.

\subsubsection{Term Occurrences in Papers}

\begin{figure*}
    \centering
    \includegraphics[width=\linewidth]{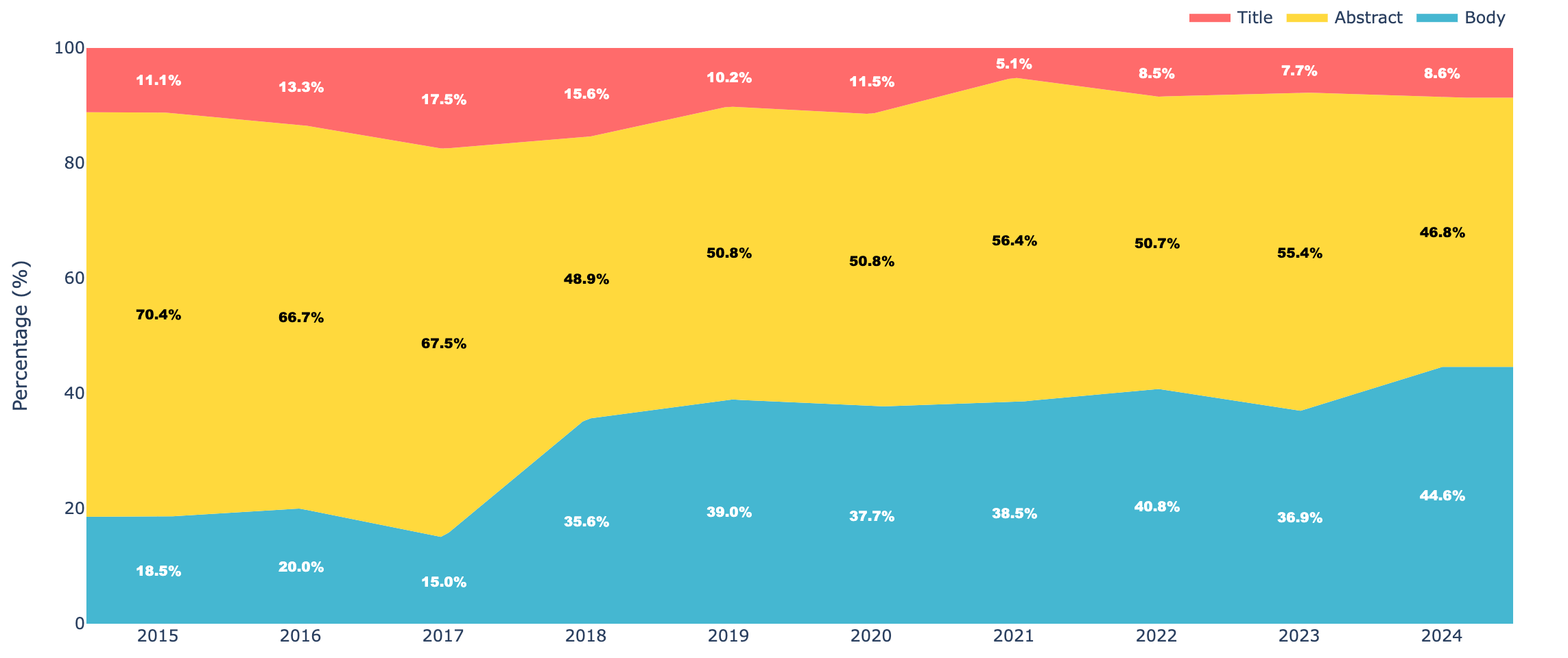}
    \caption{The position of ``framework'' in the reviewed papers (2015–2024).}
    \label{fig:position}
    \Description{The figure is a stacked area chart showing where the term framework most often appears in CHI papers from 2015 to 2024—specifically, whether it is in the title, abstract, or body of the paper. The vertical axis represents percentages, from 0 to 100, and the horizontal axis represents publication years. Three colored bands indicate the proportions: red for titles, yellow for abstracts, and blue for bodies.
    Across the decade, the majority of mentions appear in the abstract, though this share gradually decreases. In 2015, about 70 percent of papers with frameworks mention the term in the abstract, compared to 18 percent in the body and 11 percent in the title. By 2018, abstract mentions had dropped to 49 percent, while body mentions rose sharply to 36 percent. This upward trend in the body continues, reaching 45 percent by 2024. The share of title mentions remains relatively small throughout the period, fluctuating between 5 and 18 percent, with no strong upward or downward trend.
    Overall, the figure shows that while abstracts remain the most common place where frameworks are referenced, the body of papers has become increasingly important over time, narrowing the gap. Titles, meanwhile, consistently account for only a small minority of mentions.}
\end{figure*}

We categorized the corpus into three mutually exclusive groups based on the most prominent position of the term ``framework''. 
We used a hierarchical approach: 
Papers with the term in the Title were assigned first. 
From the remainder, those with the term in the Abstract were assigned. 
All other papers, where the term appeared only in the main text, were categorized as Body.

%We particularly looked at the relationship with the term ``framework'' appear position and their engagement. 
% The detailed table is showed in Appendix~\ref{app:comprehensiveanalysis}. 
%Our analysis shows that the position of the term ``framework'' within a paper is not random but strongly signals the role it plays.

Across 10 years, 61 papers used  ``framework'' in their title. 
When ``framework'' appears in the title, it is an almost unequivocal signal that the paper's core contribution is to create a new framework. 
This trend is robust and consistent; for example, in 2019, 2021, and 2024, 100\% of papers with ``framework'' in the title were coded as \Code{Create}. 
%This suggests that authors use the title to clearly declare that a framework is their work's primary contribution.

The largest group, with 330 papers, consisted of those in which the term appeared in the abstract but not in the title.
Its engagement type is distribution is less skewed towards \Code{Create} than the title.
It typically reflects a mix of all engagement types, \revtwo{largely mirroring the overall distribution observed across the full corpus.}

There are 224 papers used term only in the papers' body,
the dynamic of engagement type shifts significantly in the paper's body compare with title and abstract. 
As one might expect, this is the primary site for supporting activities, where the proportions of \Code{Use} and \Code{Mention} increase substantially as frameworks are applied as conceptual tools or cited to provide contextual information. 
However, even when the term ``framework'' appeared only within the body of the paper, an average of 36.9\% of such cases still represented \Code{Create} framework contributions.

Finally,  the proportional distribution of the term's position has remained remarkably stable since 2018, as shown in Figure \ref{fig:position}. 
This stability likely reflects the emergence and solidification of scholarly norms within the CHI community. As the concept of a ``framework'' became more common, authors and reviewers implicitly developed a shared understanding of how to we interact with the framework contribution.

\subsubsection{\rev{Engagement Type and Contribution Type}}

\begin{figure*}
    \centering
    \includegraphics[width=\linewidth]{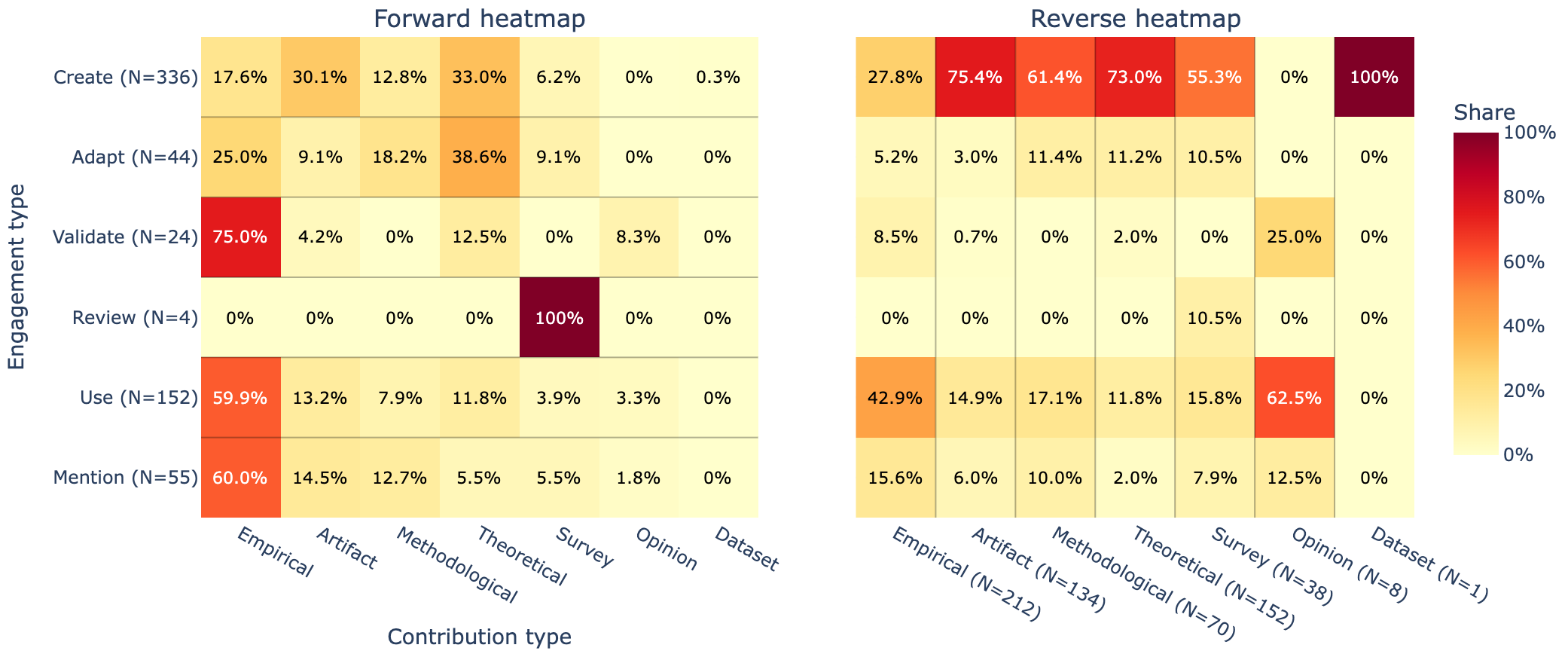}
    \caption{
    \revtwo{Dual-direction cross-tabulation of engagement and contribution types. The left heatmap shows, for each engagement type (rows), the percentage distribution of contribution types (columns). The right heatmap shows, for each contribution type (columns), the percentage distribution of engagement types (rows)}
    % \rev{
    % A balloon plot visualizing the epistemic relationship between engagement types (rows) and contribution types (columns). Color intensity and the upper-left numeric label represent the forward percentage (the proportion of a specific contribution type produced by each engagement). Circle size and the down-right numeric label represents the reverse percentage (the proportion of an engagement type accounting for a specific contribution).}
    }
    \Description{The figure is a bubble chart showing how different types of contributions relate to five types of engagement: Create, Adapt, Validate, Review, and Use. Contribution types appear along the horizontal axis, and engagement types along the vertical axis. Each bubble represents the share of a contribution type within a given engagement, with bubble size indicating how common that engagement is for that contribution type, and color intensity indicating the contribution share within the engagement. The chart highlights that empirical contributions dominate validation and use, while theoretical and artifact contributions are more evenly distributed across creation and adaptation.  A color bar on the right provides the scale for contribution share.}
    \label{fig:twowayheat}
\end{figure*}

\rev{
To understand the relationship between how researchers engage with frameworks and the primary contribution type of their paper, we conducted a dual-directional cross-analysis of Engagement Types against coded Wobbrock and Kientz's Contribution Types.
% Figure~\ref{fig:twowayheat} visualizes this relationship: the \textbf{color intensity} represents the composition of contributions within each engagement type (forward \%), while the \textbf{circle size} the composition of engagement within each contribution type (reverse \%).
}
\revtwo{Figure~\ref{fig:twowayheat} visualizes this relationship: 
The left forward heatmap represents the percentage of contributions per engagement type, whereas the right heatmap reverses this view to show the share of engagement types for each contribution type.}

%IMPORTANT POINT
\rev{
From a forward perspective, we observe that \Code{Create} papers are distributed relatively evenly across the first five contribution types.
Although \textit{Theoretical} contributions remain the most frequent (33\%), this appears different to Wobbrock and Kientz's classification, which classify frameworks as a theoretical contribution.
We propose two reasons for this: First, the framework may merely be a byproduct of the paper and thus not considered the primary contribution. 
Second, authors may subjectively claim or label a certain form of knowledge as a ``framework,'' regardless of whether it really qualify as a framework.
In contrast, \Code{Validate} papers are predominantly associated with \textit{Empirical} contributions (75\%). This is because validation inherently involves the testing of a framework's applicability or properties through new empirical data. 
}

\revtwo{Viewing the data in reverse heatmap} \rev{reveals that a \textit{Survey} contribution does not necessarily correspond to a \Code{Review} paper. Instead, 55.3\% of papers with a Survey contribution actually lead to the \Code{Creation} of a new framework. 
This aligns with our analysis of construction methods in later sections, which shows that a significant body of work employs literature reviews as a generative method to synthesize and construct new frameworks.
}

\rev{
Both forward and reverse analyses highlight a tight coupling between \Code{Use} engagement and \textit{Empirical} contributions. In \Code{Use} papers, 59.9\% result in an empirical contribution; conversely, within the corpus of Empirical contributions, 42.9\% of papers involve the use of a framework. This indicates that a main utility of frameworks in HCI is as analytical tools for interpreting empirical data.
}

% we can see that \Code{Create} papers distributed pretty equally on first five different paper contributions. Although primary theoretical contribution still the most 33\%. 
% This slightly contradicts Wobbrock's classification that the framework constitutes a theoretical contribution. 
% There are two reasons for this: first, the framework may merely be a byproduct of the paper and thus not considered the primary contribution. Second, the authors arbitrarily claimed a certain form of knowledge as the framework.

% \Code{Validate} papers usually come with empirical contribution (75\%). This is because...

% Reversely, 
% we can see survey contribution doesn't necessarily be a \Code{Review} paper. Instead, 55\% of survey contribution papers lead to \Code{Create} frameworks. This also align with our analysis about how authors construct frameworks in later section, pretty a big body of work use literature review to form their framework. 

% Both forwardly and reversely, we can see the close relationship between \Code{Use} engagement and the empirical contribution type. In \Code{Use} papers, 60\% of them lead to a empirical contribution. and In empirical contributions, 43\% paper they use framework. This indicate...

\subsubsection{\rev{Framework Usage Across Domain}}

\begin{figure*}
    \centering
    \includegraphics[width=0.95\linewidth]{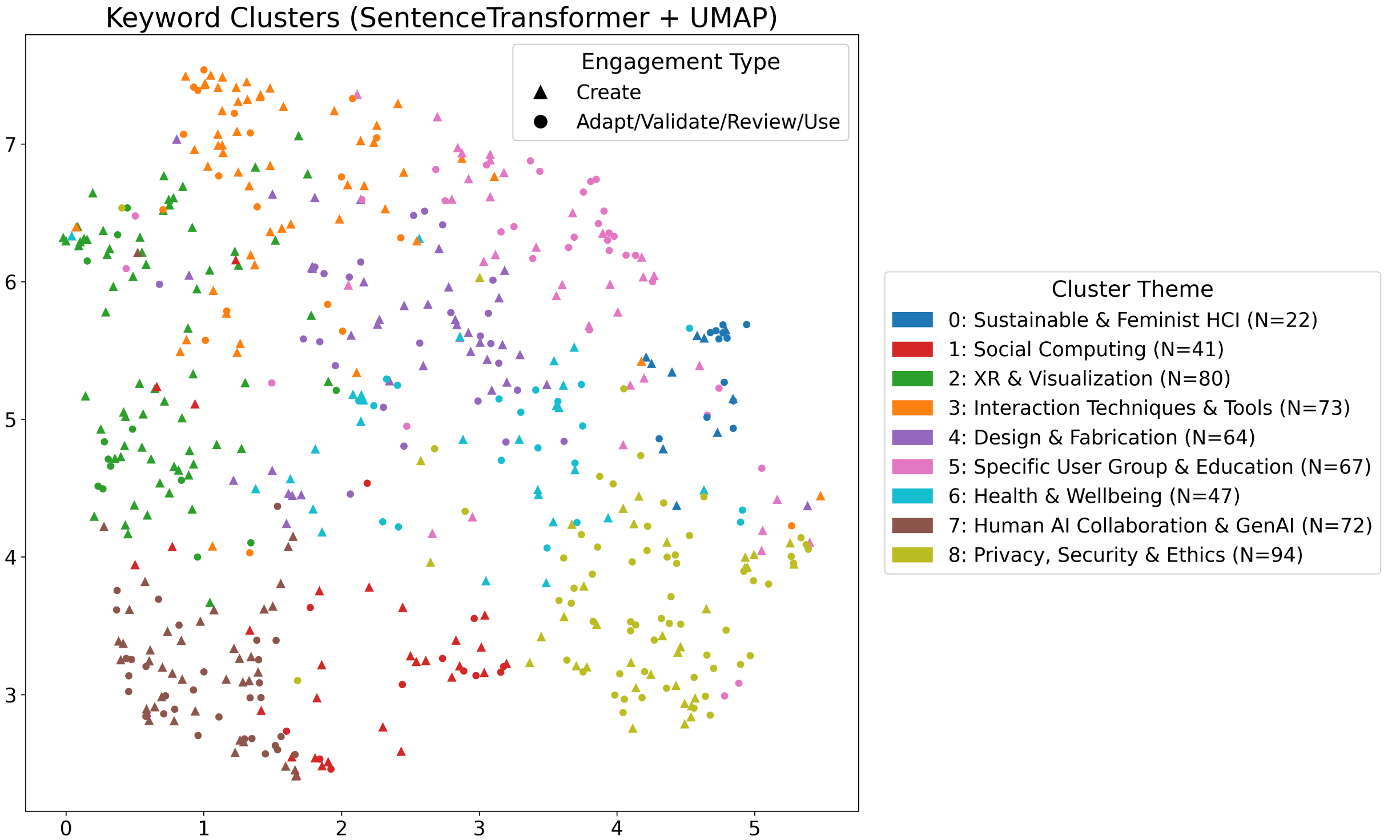}
    \caption{\rev{A UMAP projection of the semantic space of framework-term-engaged papers (N=560) based on their keywords. Papers were embedded using Sentence-Transformer and clustered into nine thematic domains (represented by colors). Markers distinguish the engagement type. }}
    \label{fig:domaincluster}
    \Description{The figure shows a scatterplot of keyword embeddings projected with UMAP, where each point represents a paper and is colored according to its thematic cluster. Nine clusters are displayed, covering topics such as sustainable HCI, social computing, XR and visualization, interaction techniques, design and fabrication, specific user groups, health and wellbeing, human–AI collaboration, and privacy and ethics. Points also vary in shape to indicate engagement type, with triangles representing papers that create frameworks and circles representing papers that adapt, validate, review, or use them. The plot illustrates how papers group together by thematic similarity and how engagement types are distributed across clusters.}
\end{figure*}

\rev{
To understand the domain contexts in which frameworks are created and deployed, we moved beyond aggregate counts to examine the specific subfields of HCI where these epistemic activities occur. 
We  focused on \Code{Create}, \Code{Adapt}, \Code{Validate}, \Code{Review}, and \Code{Use} that have deeper engagement with the frameworks and conduct analysis on this sub-corpus (N = 560). The details of the analysis in in \autoref{app:clustering} and the key results are shown in \autoref{fig:domaincluster}.
}

\rev{
Cluster 8 (Privacy, Security, \& Ethics) contains the highest number of papers ($N=94$). This prominence aligns with our earlier observation of frequently used frameworks (Table~\ref{tab:frameworks used}), where established theoretical lenses for privacy (e.g., Contextual Integrity~\cite{nissenbaum2004privacy}) and ethics (e.g., Tronto’s Care Ethics~\cite{tronto2020moral}) appeared repeatedly.
}

\rev{
In terms of spatial distribution, Cluster 0 (Sustainable \& Feminist HCI) exhibits the highest density in the projected latent space. This concentration indicates a high degree of semantic cohesion within this community. Unlike more diffuse technical clusters where terminology might vary by application, papers in this cluster tend to share a tight, specialized vocabulary. 
}

\begin{figure*}
    \centering
    \includegraphics[width=\linewidth]{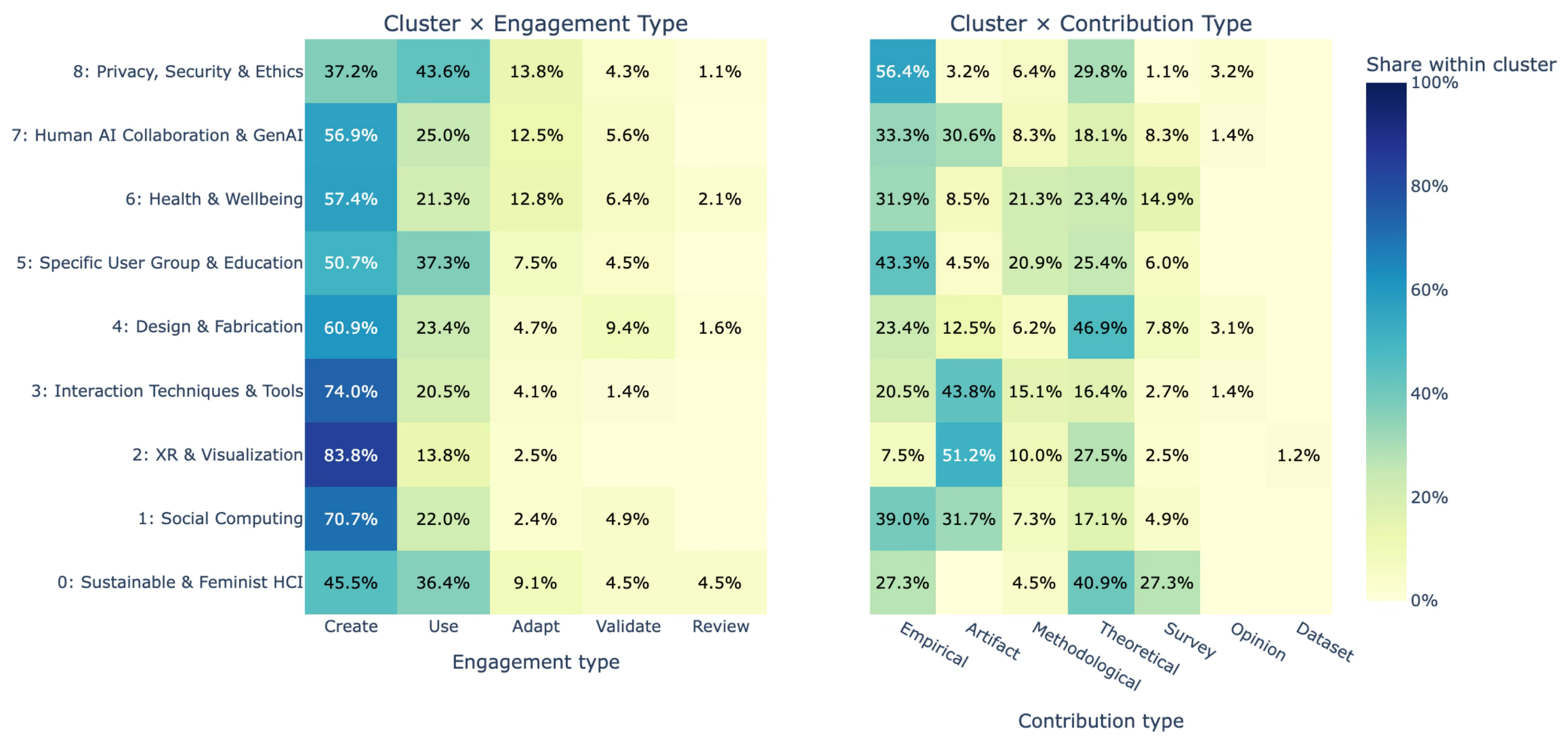}
    \caption{\rev{Distribution of engagement and contribution types across keyword clusters. The left panel shows the percentage of papers in each cluster performing specific engagement activities (e.g., Create, Use). The right panel shows the primary contribution type of those papers (e.g., Empirical, Artifact). Darker shades indicate a higher proportion of papers within that specific cluster.}}
    \label{fig:heatmap}
    \Description{The figure contains two heatmaps summarizing how thematic clusters relate to engagement types and contribution types. The left heatmap maps nine clusters to five engagement types, showing the percentage share of each engagement within a cluster. The right heatmap maps the same nine clusters to six contribution types, again showing the percentage share within each cluster. Darker colors indicate higher shares, with a color bar on the right providing the scale from low to high values. Together, the heatmaps show distinct patterns across clusters, such as higher rates of creation in areas like XR and interaction techniques, and varying distributions of empirical, artifact, theoretical, and survey contributions across clusters.}
\end{figure*}

\rev{To further characterize the epistemic nature of these domains, we examined the distribution of \textbf{Engagement Types} and \textbf{Contribution Types} within each cluster (Figure~\ref{fig:heatmap}).}

\rev{Clusters rooted in technical innovation, specifically Cluster 2 (XR \& Visualization) and Cluster 3 (Interaction Techniques \& Tools), exhibit the highest rates of \Code{Create} engagement (83.8\% and 74.0\%, respectively). Correspondingly, their contribution types are heavily skewed towards Artifacts (51.2\% for XR; 43.8\% for Interaction).
}

\rev{
In contrast, domains dealing with complex social norms or specific user populations --- such as Cluster 8 (Privacy, Security \& Ethics) and Cluster 5 (Specific User Group \& Education) --- show a much higher prevalence of \Code{Use} (43.6\% and 37.3\%, respectively). The primary contribution in these clusters is Empirical (56.4\% for Privacy; 43.3\% for Education).}

%%%%%%%%%%%%%%%%%%%%%
\subsection{Understanding Framework Contributions Through A Functional Typology (RQ2)}

Our analysis of RQ1 showed that new framework contributions happen as  \Code{Create} and \Code{Adapt} papers, which together account for over 60\% of our corpus. 
To answer RQ2, we next consider the substance of these framework contributions. 
We first examine how authors themselves label their work and then use a typology of functions to clarify the diverse forms and functions these contributions take.

% Our landscape analysis in the previous section revealed that the discourse around frameworks is dominated by two primary activities: the \Code{Create} engagement in proposing new contributions and the \Code{Use} engagement in applying existing ones. 
% The Create and Adapt types, which together account for over 60\% of our corpus, represent the primary sites where new conceptual contributions are being offered to the HCI community.

% We next look at the substance of the framework contributions, our analysis now zooms in on this specific subset of papers. 

% In this section, we address RQ2 by deconstructing these created and adapted frameworks. We first examine how authors themselves label their work and then propose a functional typology to bring clarity to the diverse forms and functions these contributions take.

\subsubsection{How Author Label Their Framework}

\begin{table*}[t]
\caption{The Nine Most Common Self-Reported Framework Types in the Sub-corpus (N = 380)} 
\label{tab:labeltype}
    \centering
    {\scriptsize
    \begin{tabularx}{\textwidth}{p{2.5cm} p{0.5cm} X}
    \toprule
    \textbf{Self-labeled Type} &\multicolumn{1}{c}{\textbf{Count}} & \multicolumn{1}{c}{\textbf{Illustrative Quotes}} \\
    \midrule

Conceptual Framework
& 45 &
``Building on the results, we propose a conceptual framework for VR onboarding and discuss...''~\cite{10.1145/3544548.3581211}

``These competencies and design considerations are organized in a conceptual framework thematically derived from the literature''~\cite{10.1145/3313831.3376727}
    % \item ``This work contributes a conceptual framework that considers privacy choice as a user-centered process as well as a taxonomy for ...''~\cite{10.1145/3411764.3445148}
\\
\addlinespace
Design Framework 
& 38 &
    % \item ``The findings include a general design framework based on an iterative participatory design focusing on preferred activities...''~\cite{10.1145/3613904.3642595}
``... we discuss the stakes of these choices, suggest future research directions, and propose an emerging design framework for shaping pro-social behavior in VR''~\cite{10.1145/3290605.3300794}

``Inspect, Embody, Invent: A Design Framework for Music Learning and Beyond'' the title of ~\cite{10.1145/2858036.2858577}
\\
\addlinespace

Theoretical Framework 
& 20 &
    % \item ``These investigations form the theoretical framework for the subsequent analysis of five digital body game examples.''~\cite{10.1145/3411764.3445622}
``Specifically, this study contributes to the following areas: (1) offering a theoretical framework that can be used to guide the design and evaluation of learning with technologies...''~\cite{10.1145/3544548.3580913}

``Yes: Affirmative Consent as a Theoretical Framework for Understanding and Imagining Social Platforms'' the title of~\cite{10.1145/3411764.3445778}
\\
\addlinespace

Analytic(al) Framework 
& 15 &
    % \item ``we build on prior studies to propose an analytical framework of personal informatics for mental wellness.''~\cite{10.1145/3173574.3174146}
``We first show how our taxonomy can serve as an analytical framework for video navigation systems''~\cite{10.1145/3544548.3581126}

``We begin by proposing an analytical framework to highlight the importance of calibrated human self-confidence''~\cite{10.1145/3613904.3642671}
\\
\addlinespace

Interaction Framework 
& 11 &
``Our findings resulted in a
human-notification interaction framework comprised of 12 unique
motivations frequently associated with three activity timings for
interacting with notifications, including...''~\cite{10.1145/3544548.3581146}
    
    % \item ``A new interaction framework for human-AI collaboration in creative tasks that build on the premise that LLMs should allow users to explore a space of possible responses, rather than giving a single data point in response to user input.''~\cite{10.1145/3613904.3642400}
``A Visual Interaction Framework for Dimensionality Reduction Based Data Exploration'' the title of ~\cite{10.1145/3173574.3174209}
\\
\addlinespace

Evaluation Framework 
& 10 &

``Hence, to facilitate research and applications that use touch biometrics, we contribute: ... 3) an evaluation framework to estimate the expected amount of user-revealing information in touch interactions with given interfaces''~\cite{10.1145/2858036.2858165}

``we present a Diversity Prompting Evaluation Framework consolidating metrics from several research disciplines to analyze along ...''~\cite{10.1145/3411764.3445782}
    % \item ``We develop our evaluation framework based on the adult-child communication literature that ...''~\cite{10.1145/3411764.3445271}
\\
\addlinespace

Methodological Framework &
6 &
``This paper offers a theoretical and methodological framework of `techno-aesthetic encounters' that supports nonlinear and art-based modes of inquiry in HCI and the broader STEM fields''~\cite{10.1145/3491102.3517506}
    % \item ``Thus we present the Design Study "Lite" Methodology, a novel framework for implementing design studies with novice students in 14 weeks''~\cite{10.1145/3313831.3376829}
    
``Towards a Non-Ideal Methodological Framework for Responsible ML''~\cite{10.1145/3613904.3642501}
\\
\addlinespace

Computational Framework 
& 4 & 
    % \item ``We develop a novel computational framework to address these RQs and explore differences in linguistic norms.''~\cite{10.1145/3173574.3174240}
``A computational framework for quantifying the defectiveness of interview chatbots. This framework comprehensively evaluates...''~\cite{10.1145/3411764.3445569}

``Fast-Forward Reality first introduces a computational framework that supports generating high-quality test cases tailored to ...''~\cite{10.1145/3613904.3642158}
\\
\addlinespace

Technical Framework 
& 4 &
``... we introduce model sketching: a technical framework for iteratively and rapidly authoring functional approximations of a machine learning model’s decision-making logic''~\cite{10.1145/3544548.3581290}

``Our paper focuses on the comprehensive technical framework to detecting SIIDs (situationally induced impairments and disabilities)...''~\cite{10.1145/3613904.3642065}
\\
\addlinespace

\bottomrule
\end{tabularx}
}
\end{table*}

Table \ref{tab:labeltype} shows that authors use a diverse but recurring set of words to describe their framework contributions. 

According to our count, the nine most common self-labeled types are: conceptual (45), design (38), theoretical (20), analytical (15), interaction (11), evaluation (10), methodological (6), and computational and technical frameworks (4 each).

% Table \ref{tab:claimed_type} demonstrates that authors use a diverse but recurring set of labels to describe their framework contributions. 
% Across our corpus, nine types are most frequently observed: \textit{conceptual, design, theoretical, evaluation, methodological, analytical, computational} and \textit{technical frameworks}. 

% Based on our qualitative observation, conceptual frameworks are the most dominant, 
% followed by \textit{design frameworks} and \textit{theoretical frameworks}, 
% while the remaining types appear with lower frequency. 
% We cannot provide precise statistics here, as there is no standardized rhetorical system for labeling frameworks, and many papers do not explicitly assign a type at all. 
% Instead, we report qualitative observational frequencies, which should be interpreted as indicators of relative salience rather than exact counts.

The prominence of \textit{conceptual and theoretical frameworks} is unsurprising, as earlier work show their role in structuring and explaining research~\cite{imenda2014there,ravitch2016reason, varpio2020distinctions}.
However, more distinctive to HCI is the frequent appearance of \textit{design frameworks}, which reflects the field’s interest in actionable design support and the need to provide generative guidance for design-oriented research.

We also found several cases where the authors assigned multiple labels to a single framework. 
For example, a framework for measuring the impact of the introduction of AI systems in decision settings~\cite{10.1145/3544548.3581095} was described simultaneously as a \textit{conceptual} and a \textit{methodological} framework. Similarly, both~\cite{10.1145/3491102.3517506} and~\cite{10.1145/2702123.2702176} characterized their contributions as \textit{theoretical} as well as \textit{methodological} frameworks. 
In other cases, the authors position an existing theory as a \textit{theoretical framework} but later suggested that it could also serve as a \textit{design framework} --- for example, the frameworks of the feminist theory of affirmative consent~\cite{10.1145/3411764.3445778} and the Mutual Theory of Mind~\cite{10.1145/3411764.3445645}. 
These instances illustrate that authors do not always apply a single, still label, but instead mobilize multiple descriptors to signal different facets of their contributions.

However, when we attempted to use these labels as the basis for a typology of frameworks, we encountered substantial overlap and ambiguity. 
The same term is often applied to contributions with very different goals and structures. 
For example, a framework labeled as \textit{design} may in practice serve a primarily \textit{conceptual} purpose \cite[e.g.,][]{10.1145/3290605.3300794}; a \textit{theoretical} framework may be mobilized to guide \textit{design} \cite[e.g.,][]{10.1145/3544548.3580913}; and a \textit{computational} framework may function as both an \textit{evaluation} and an \textit{analytical} tool \cite[e.g.,][]{10.1145/3411764.3445569}. 
This inconsistent lexicon makes it difficult to infer a function from the author’s label alone. 

\subsubsection{Understanding Contribution Through Function}

Inspired by the synthesis of framework functions in the information systems literature by Schwarz et al.~\cite{Schwarz07understanding}, and by revisiting Bederson and Shneiderman’s~\cite{bederson2003craft} classic classification of theory types in HCI, we change perspective to focus on what frameworks \textit{do} rather than what they are called, addressing RQ2.

Bederson and Shneiderman distinguished five ways that theories may contribute: \textit{descriptive, explanatory, predictive, prescriptive}, and \textit{generative}. Since HCI frameworks are often considered a form of theoretical contribution~\cite{wobbrock16contribution}, we draw on this typology to map the functions that frameworks serve in our corpus.
In this typology, \textit{descriptive} frameworks serve to organize and categorize phenomena, providing researchers with shared vocabulary and structures to characterize interactional contexts. 
\textit{Explanatory} frameworks go further by articulating causal relationships, helping scholars and practitioners understand why certain behaviors or outcomes emerge in interaction. 
\textit{Predictive} frameworks enable anticipation of future outcomes or user behaviors, often supplying models that can be applied across contexts to forecast performance or usability. 
\textit{Prescriptive} frameworks embody actionable guidance, offering heuristics, design principles, or rules that practitioners can directly apply in building systems. 
\textit{Generative} frameworks open up new conceptual spaces, metaphors, or perspectives that inspire innovative designs.

\begin{table*}[h]
\centering
\caption{Mapping of Five Types of Contribution to Framework}
\label{tab:function_mapping}
\begin{tabular}{p{3cm} p{7cm} p{4cm}}
\toprule
\textbf{Contribution Type} & \textbf{Associated Functions} & \textbf{Example} \\
\midrule

Descriptive & 
Organize and categorize phenomena; provide shared vocabulary and structures for interactional contexts. & 
~\cite{10.1145/3290605.3300480}%high-level categories that describe the design space for social drones through lituerature review. in paper: Rather, our framework means to be descriptive rather than prescriptive
~\cite{10.1145/3544548.3581211}\textsuperscript{\ensuremath{\ast}} %performed 21 VR tutorial ergonomic reviews and 15 interviews with VR experts with experience in VR onboarding. it also support generate new design
~\cite{10.1145/3173574.3173714} %visual interaction cues framework summarized from 49 popular contemporary video games
~\cite{10.1145/3290605.3300658}\textsuperscript{\ensuremath{\diamond}},%demonstrated by its explanatory power and its ability to generate concrete design guidance
~\cite{10.1145/3290605.3300767} %characterize MR applications in terms of the number of environments, number of users, level of immersion, level of virtuality, degree of interaction, input, and output with literature review and expert interview
~\cite{10.1145/3411764.3445088}\textsuperscript{\ensuremath{\star}} %creation of a new, more granular and composable vocabulary for characterizing a complex sociotechnical space. This act of re-describing the stakeholders and their needs is what enables the framework's other powers
% ~\cite{10.1145/3411764.3445782} % Directed Diversity: Leveraging Language Embedding Distances for Collective Creativity in Crowd Ideation: The authors note, "our proposed framework is descriptive to curate many useful metrics, but not prescriptive to recommend best metrics"

\\
\addlinespace

Explanatory & 
Articulate causal relationships; explain why particular behaviors or outcomes emerge. & 

~\cite{10.1145/2702123.2702232}% It primarily explains how and why people's everyday practices—and the energy consumption that results from them—are shaped by a complex interplay of factors, rather than just by individual choice and motivation. quote: COWOP can be seen as an exploratory and explanatory conceptual framework—and not as a prescriptive tool—to support HCI researchers and designers in identifying, analyzing, and understanding practices to support system design.
~\cite{10.1145/3411764.3445778}\textsuperscript{\ensuremath{\ddag}} %Affirmative Consent as a Theoretical Framework quote: this paper argues that affirmative consent is both an explanatory and generative theoretical framework. 
~\cite{10.1145/3290605.3300297}\textsuperscript{\ensuremath{\dag}} %explain how and why trust is formed in social groups. build statistical models to predict a user's trust score based on a wide range of variables
~\cite{10.1145/3290605.3300658}\textsuperscript{\ensuremath{\diamond}} %conducted classroom observations and teacher interviews to identify and categorize how teachers use the ST Math game
\\
\addlinespace

Predictive & 
Enable anticipation of future outcomes or user behaviors; apply models across contexts to forecast performance or usability. & 
~\cite{10.1145/3290605.3300451} %the framework is the architectural blueprint for that system. Its success is measured almost entirely by its ability to accurately predict personality traits from physiological data
~\cite{10.1145/3411764.3445563} %ML predictive model
~\cite{10.1145/2858036.2858057} % generate testable predictions. The authors use the descriptive axes to formulate ten specific hypotheses (H1-H10) that predict how text legibility will be affected by different shape properties and text mappings. structure the scientific inquiry by generating a series of testable predictions (hypotheses). The framework is the engine that drives the entire experimental methodology of the paper; the four studies exist solely to test the validity of its predictions.
~\cite{10.1145/3290605.3300297}\textsuperscript{\ensuremath{\dag}}
\\
\addlinespace

Prescriptive & 
Provide actionable guidance, heuristics, design rules, evaluation metrics, or methodological procedures to inform practice.
& 

~\cite{10.1145/3313831.3376733} %By empirically ranking the effectiveness of different sensory channels (scent type, scent intensity, airflow, temperature) for different data types (nominal, ordinal, quantitative), the work provides direct, actionable guidance for designers.
~\cite{10.1145/3613904.3642781} % It directly addresses the question of what designers should do to create ethical designs, rather than just what to avoid. It prescribes that designers should align their work with positive, "standard concepts" documented in a "concept catalog". 
~\cite{10.1145/3411764.3445341}\textsuperscript{\ensuremath{\circ}} %Designing Civic Technology with Trust. offering a clear, "stepwise order" for designers to follow when designing with trust. The authors explicitly call it a "generative frame" intended to keep trust "central in ideation". 
~\cite{10.1145/3290605.3300240} %four-stage process to guide for designers and developers to create accessible interactive tools for children with disabilities.
~\cite{10.1145/3290605.3300658}\textsuperscript{\ensuremath{\diamond}},%demonstrated by its explanatory power and its ability to generate concrete design guidance

 \\
\addlinespace

Generative & 
Open up new conceptual spaces, metaphors, or perspectives that inspire innovative designs and novel lines of inquiry. 
& 

~\cite{10.1145/3411764.3445341}\textsuperscript{\ensuremath{\circ}}%Its third and final step uses four "sensitizing concepts" (e.g., historicizing engagement, focusing on experience) to guide a "generative design process" where specific system features are created.
~\cite{10.1145/3544548.3581211}\textsuperscript{\ensuremath{\ast}}
~\cite{10.1145/3411764.3445778}\textsuperscript{\ensuremath{\ddag}}
~\cite{10.1145/3491102.3502087} % scalably generate a large and diverse set of prompts. 
~\cite{10.1145/3411764.3445088}\textsuperscript{\ensuremath{\star}} % The framework is explicitly designed to be generative, helping researchers "envision new or previously unexplored points in the problem space". The authors demonstrate how its vocabulary can be used to Generate new personas, Generate new persona-need combinations, and new designs. they also evaluated desctiptive and generative power
\\
\bottomrule

\end{tabular}

\caption*{\rev{\textit{Note: Same symbols following each example (e.g., \ensuremath{\circ}, \ensuremath{\star}) denote that they are the same paper}}}
\end{table*}

% 5 design frameworks maping to different funtional frameworks 
% 10.1145/3491102.3517699 Descriptive
% 10.1145/3313831.3376460 Prescriptive
% 3173574.3173817 Descriptive
% 10.1145/3613904.3642455 descriptive
% 

Table~\ref{tab:function_mapping} shows how frameworks in our corpus align with the five functional types. 
Notably, authors are already spontaneously drawing on this typology when naming their frameworks. 
For instance, some papers explicitly call their contributions \textit{``descriptive framework''}~\cite{10.1145/3290605.3300480, 10.1145/3173574.3173714} \textit{``explanatory frameworks''}~\cite{10.1145/2702123.2702232} or even \textit{``generative framework with prescriptive elements''}~\cite{10.1145/3411764.3445341}, directly echoing the functional language of this classification. 
In this sense, our functional mapping does not impose an external scheme, but instead reflects the rhetorical practices that authors themselves adopt.

At the same time, it is clear that a single framework can embody more than one contribution type. 
For example,
The VR Onboarding Framework~\cite{10.1145/3544548.3581211} is  \textit{descriptive}, organizing the space of onboarding practices to provide categories and vocabulary for understanding onboarding processes. Meanwhile, it demonstrates generative power by opening up new design opportunities. 
The framework for trust in social groups~\cite{10.1145/3290605.3300297} is presented as an \textit{explanatory} account, showing how and why trust is formed within groups through the interplay of social factors. 
Yet, the framework is also \textit{predictive}, as it builds statistical models to forecast a user’s trust score from a wide range of variables.
Similarly, the framework on designing civic technology with trust~\cite{10.1145/3411764.3445341}, illustrates both generative and prescriptive roles. 
It generates new conceptual lenses through sensitizing concepts that inspire design ideas, while simultaneously offering prescriptive guidance as  a stepwise process for practitioners to follow when designing with trust.

Together, these observations suggest that the functional types of framework are their defining characteristics. And this resonates with how the HCI community already conceives of and communicates its framework contributions.

% It is also helpful for us to discover their common patterns and key components. 

\subsubsection{Key Constructs and Components Among Contribution Type}

\begin{figure*}
    \centering
    \includegraphics[width=\linewidth]{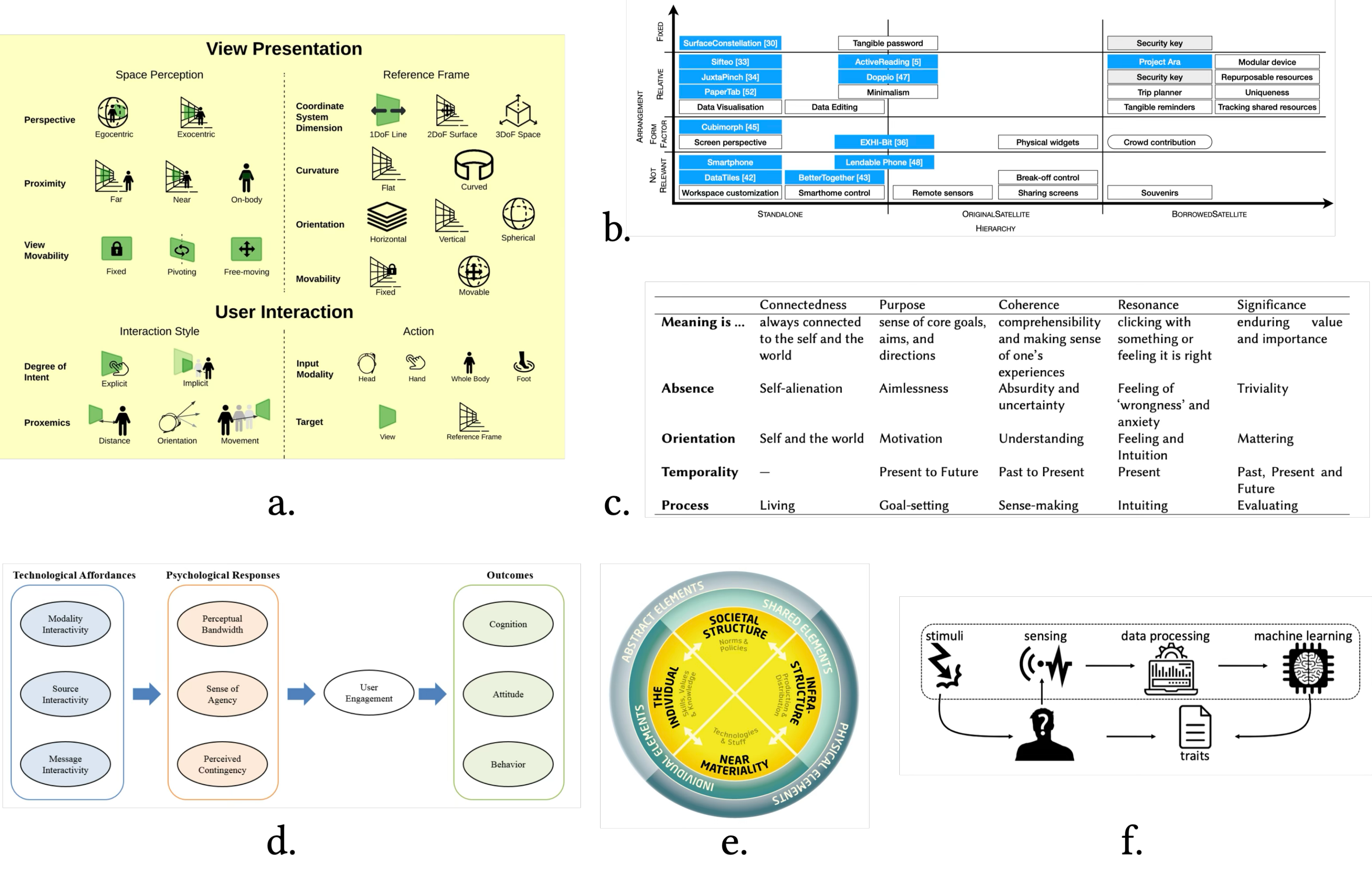}
    \caption{Typical Visualization Tools for Frameworks and Their Core Components. 
    (a) is a illustration for visualization view management for 3D space ~\cite{10.1145/3544548.3580827}. 
    (b) is a visual matrix for PickCells design space~\cite{10.1145/3290605.3300503}. 
    (c) is an table for describing five components of meaning~\cite{mekler19meaning}. 
    (d) is a visualization of interaction effects of the different types of interactivity~\cite{10.1145/3491102.3502207}. 
    (e) is a wheel graph illustrates how practice theory mediates different aspects of practice in sustainable HCI ~\cite{10.1145/2702123.2702232}. 
    (f) is a flow diagram for personality detection ~\cite{10.1145/3290605.3300451}.}
    \label{fig:visualexample}
    \Description{The figure provides an overview of typical visualization tools used to present frameworks and their core components. Subfigures (a), (b), and (c) illustrate list- or matrix-style presentations, such as a 3D view management illustration, a design space matrix, and a table describing conceptual components. Subfigures (d), (e), and (f) use arrows or flow to indicate relationships, causality, or processes, including interaction effects, mediating roles in practice theory, and a flow diagram for personality detection}
\end{figure*}

%We now move from analyzing what frameworks do to what they are made of. 
This section synthesizes the data we extracted about core conceptual components and presentational forms of frameworks in our corpus. 
We use our five functional types as an analytical lens and find that their structure and components converge across types. 
% Our analysis reveals distinct patterns in the anatomy of frameworks depending on the role they are designed to play.

Frameworks whose primary function is to be \textit{descriptive} are typically built from components that help organize and classify a domain. 
The most common conceptual tools we observed were taxonomies \cite[e.g.,][]{10.1145/3290605.3300733, 10.1145/3313831.3376498},
% factors\cite[e.g.,][]{10.1145/3173574.3173732, 10.1145/3491102.3501953}, 
sets of categories \cite[e.g.,][]{10.1145/3173574.3173818, 10.1145/3290605.3300624}, 
and dimensions \cite[e.g.,][]{mekler19meaning, 10.1145/3290605.3300767}. 
These components provide a shared vocabulary and structure for a given topic. 
Visually, they are often presented as tables, matrices, or diagrams that map the conceptual space (as in Fig.~\ref{fig:visualexample} a.b.c.).

Frameworks designed to be \textit{explanatory} or \textit{predictive} tend to be composed of tools that articulate relationships and causality. 
Their core components are often interrelated factors or variables \cite[e.g.,][]{10.1145/3290605.3300297}, and a set of propositions or hypotheses that define the connections between them \cite[e.g.,][]{10.1145/3491102.3502207, 10.1145/2858036.2858057}. 
For \textit{predictive} frameworks specifically, this often includes statistical or computational models \cite[e.g.,][]{10.1145/3290605.3300451}. 
These are typically visualized through flow diagrams illustrating causal links or architectural diagrams showing system components and procedures (as in Fig.~\ref{fig:visualexample} d.e.f.). 

When a framework's function is \textit{prescriptive}, its components are almost always a form of actionable guidance. 
We found that these frameworks are most commonly built from design principles, heuristics, guidelines, evaluation metrics, or step-by-step processes \cite[e.g.,][]{10.1145/3290605.3300773}. 
These tools are meant to be applied directly by practitioners. 
Consequently, they are often presented in easily digestible formats such as numbered lists, checklists, and process diagrams. 

Finally, \textit{generative} frameworks are constructed from more open-ended and inspirational components. 
Rather than providing direct answers, they offer tools to provoke new thinking. 
The most common components were sensitizing concepts, lenses, metaphors, and perspectives designed to help researchers and designers ``envision new or previously unexplored points in problem space''~\cite{10.1145/3411764.3445088}.
Their form is often more abstract, relying on rich narrative descriptions and illustrative examples to convey the core concepts.
Just as a framework can serve multiple functions, it can also be built from a variety of components.

\subsection{The Craft of Framework: Its Construction, Validation, and Articulation (RQ3)}

We now turn to our final research question, 
RQ3: How are these proposed frameworks constructed, validated, and articulated for other people's use?
% This section analyzes the methodological practices behind the frameworks in our corpus. 
Our analysis reveals that a framework's functional type often shapes the methods used to develop it, but that validation practices  are less consistent across all types. Reuse guidance is even more rare. 

\subsubsection{How are Frameworks Constructed?}

%TO DO, provide examples
Our analysis of the reported development methodologies shows a clear pattern: The methods used to build a framework align with its intended function.

Frameworks with a \textit{descriptive} or \textit{explanatory} function are typically constructed by synthesizing existing knowledge or analyzing new empirical data. 
One common approach is literature synthesis, in which authors systematically review a body of work to derive descriptive categories and map a research domain \cite[e.g.,][]{10.1145/3290605.3300831, 10.1145/3290605.3300767}. 
Alternatively, authors conduct qualitative empirical studies --- such as domain investigations \cite[e.g.,][]{10.1145/3544548.3581211, 10.1145/3290605.3300794} and semistructured interviews \cite[e.g.,][]{10.1145/3025453.3025560, 10.1145/3613904.3641976} --- to synthesize new concepts or explanatory relationships directly from their findings.

% \textit{Descriptive} and \textit{explanatory} frameworks are most often constructed through literature synthesis \cite[e.g.,][]{10.1145/3290605.3300831, 10.1145/3290605.3300767}, domain investigation \cite[e.g.,][]{10.1145/3544548.3581211, 10.1145/3544548.3581211}, or thematic synthesis through qualitative interview data \cite[e.g.,][]{10.1145/3025453.3025560, 10.1145/3613904.3641976}. 
% Authors typically derive descriptive categories by systematically reviewing a body of work or generate explanatory relationships through methods like thematic analysis of interviews and observations.

\textit{Predictive} frameworks were often developed through quantitative empirical work. 
Common practice involves collecting a dataset and then using statistical modeling or machine learning to build a model capable of forecasting outcomes \cite[e.g.,][]{10.1145/3411764.3445563, 10.1145/3290605.3300451}.

As \textit{prescriptive} and \textit{generative} frameworks are more oriented toward practice, their development often stems from design-related activities. 
Common methods include distilling design principles from a series of case studies, reflecting on a Research through Design (RtD) process, or synthesizing best practices from existing literature into actionable guidelines.

\subsubsection{How Frameworks are Justified?}

Although the development methods showed a relatively clear pattern, the methods used to validate the proposed frameworks were more varied and, in many cases, absent. 
Validation is the process of providing evidence for the theoretical soundness or practical utility of a framework. 
It is vital to establish trustworthiness that the framework users need to build on trust in the framework.
Our synthesis of the existing validation practices also shows their alignment with the five functional roles, by evaluating \textit{descriptive, explanatory, predictive, prescriptive, and generative} power.

%Evaluate descriptive power of framework
To evaluate a framework's \textit{descriptive} power is to assess how well its categories, dimensions, taxonomies, or vocabulary can accurately and comprehensively capture and organize a phenomenon. 
A common method for demonstrating descriptive power is to apply the framework in illustrative cases or exemplars, showing that it can successfully structure and make sense of a complex real-world example.  
For instance, \citet{10.1145/3173574.3173714} applied their visual cues framework --- originally developed from video games --- to the new domain of Augmented Reality, thereby demonstrating its ability to describe visual phenomena across contexts. 
Similarly, \citet{10.1145/3290605.3300503} on modular touchscreen interactions and \citet{10.1145/3411764.3445088} on interpretable ML explicitly included sections validating descriptive power, showing how their frameworks could position and characterize existing work within a structured space.  

Descriptive power may also be demonstrated more implicitly through the development method itself. 
For example, \citet{10.1145/3290605.3300733} proposed a framework of 23 nudge mechanisms derived from a review of 71 papers. 
Although they did not show any validation process, the comprehensiveness and systematic nature of their method already served to substantiate the descriptive adequacy of the framework.  

% To evaluate a framework's \textit{descriptive} power is to assess how well its summarized categories, dimensions, taxonomies, or vocabulary can accurately and comprehensively capture and organize a phenomenon. 
% The typical method for demonstrating descriptive power is through application in illustrative cases or exemplars where authors show that the framework can successfully structure and make sense of a complex real-world example. 

% For example, ~\cite{10.1145/3173574.3173714} used the visual cues framework developed from video games as a basis, to describe visual cues in a different target domain, Augmented Reality. 
% ~\cite{10.1145/3290605.3300503} on modular touchscreen and ~\cite{10.1145/3411764.3445088} on interpretable ML explicitly showed a section about how they validate their \textit{descriptive} power by showing how their framework could position and charaterize existing works. 

% However, the descriptive power could also implicitly demonstrate in the method, for example, ~\cite{10.1145/3290605.3300733} did not cover content for validation, they proposed a framework of 23 nudge mechanisms identified in the 71 papers. their method already covered. 

%Evaluate explanatory power of framework
Evaluating explanatory power involves testing a framework’s capacity to clarify causal relationships and account for why particular outcomes or behaviors emerge. Much like assessments of descriptive power, the most common method is the qualitative case study, where the framework is applied as an analytical lens to demonstrate that it yields a convincing explanation of real-world phenomena.
For example, the explanatory power of the COWOP framework~\cite{10.1145/2702123.2702232} was shown in its ability to reveal how everyday energy consumption practices are shaped by complex factors beyond individual choice. Similarly, \citet{10.1145/3290605.3300297} established explanatory power by demonstrating how their trust framework captured the interplay of social variables in the formation of group trust, supported by observational findings and statistical modeling.

%evaluating predictive paower
The validation of the \textit{predictive} power of the framework is the most direct form , and papers in our corpus commonly report this validation process. 
It usually involves the empirical testing of a framework's ability to accurately forecast outcomes. 
For example, \citet{10.1145/2858036.2858057} derived hypotheses about text legibility from their framework and conducted multiple studies to test these predictions, directly assessing predictive capacity. Similarly, \citet{10.1145/3290605.3300451} validated the predictive power by demonstrating that their framework could reliably forecast personality traits from physiological data.
The typical method is quantitative empirical testing, such as controlled experiments or applying a computational model to a holdout dataset to measure its predictive accuracy.
%How to use the framework~\c

%evaluating prescriptive power
Evaluating a framework’s prescriptive power means assessing whether its actionable guidance—principles, heuristics, or methods—is useful and effective for practitioners. In our corpus, we identified two primary pathways for demonstrating this.
The first is constructive validation, in which authors implement a system guided by their framework’s principles and then evaluate the resulting system’s performance and user experience, often through a user study. For example, \citet{10.1145/3290605.3300831} did not perform a standalone evaluation of their framework. Instead, they applied it to design an explainable AI tool for a medical scenario and assessed the tool in a co-design exercise with 14 clinicians.
The second pathway involves a more direct evaluation of the framework itself, using methods such as expert reviews \cite[e.g.,][]{10.1145/3290605.3300240, 10.1145/3544548.3581500} or design workshops \cite[e.g.,][]{10.1145/3290605.3300446}.

%evaluating generative power
Finally, evaluating \textit{generative} power involves assessing a framework's ability to inspire new ideas and open up new design spaces. 
This is often the most difficult function to validate formally.
As the generative framework always blends with other functional type, we did not  observe distinct validation process for this function. 
In our observation, most papers typically demonstrate this through application rather than validation.

Across all functional types, many papers did not provided  formal validation of their proposed frameworks. In such cases, authors often acknowledged this explicitly as a limitation. For example, \citet{10.1145/3173574.3173848} noted that future work is needed to establish their framework’s utility and effectiveness.
This recurring practice of flagging validation as a next step points to an implicit community norm: Even when not undertaken, validating a framework is regarded as both desirable and integral to its research lifecycle.

% For those generative frameworks with a few \textit{prescriptive} function, the common methods are application in generative design case studies, showing the novel artifacts it helped create, or through ideation workshops.

% Few paper lightly mentioned about evaluate its generative power, for instance, ~\cite{10.1145/3411764.3445088} demonstrate generative power through thought experiments and design extrapolations. 

\subsubsection{How are Frameworks Designed to be Reused?}

Beyond construction and validation, a framework’s long-term impact depends on its articulation—how it is packaged for others to understand and reuse. Our analysis revealed a spectrum of articulation practices, ranging from explicit, practice-oriented guidance to purely conceptual contributions.
At one end of the spectrum, only a few frameworks provided direct instructions to readers on how to use them. 
For instance, just two papers dedicated a section to ``how to use the framework'' \cite{10.1145/3491102.3501953, 10.1145/3491102.3517699}. 
A small number of prescriptive or predictive frameworks also offered stepwise, actionable methods \cite[e.g.,][]{10.1145/3544548.3580962} since their framework itself is usually a step-by-step instruction. 

A more common articulation strategy was to demonstrate use through applications within the paper itself, often as part of validation. For instance, the Experience of Meaning framework was applied to analyze CHI literature on meaningfulness~\cite{mekler19meaning}, while the City-Commons Framework for Citizen Sensing illustrated its application across a 10-month real-world deployment~\cite{10.1145/3025453.3025915}. In such cases, readers are left to infer generalizable steps, with guidance remaining implicit rather than standardized.

At the other end of the spectrum, many frameworks---particularly those labeled as Conceptual, Theoretical, or Analytical with primarily descriptive or explanatory functions---offered little to no guidance on how to use them, sometimes only hinting at the contexts in which they might apply. Their value lies in providing new lenses or vocabularies for thinking about a problem, while their reuse depends entirely on the cleverness  of the reader.

Finally, we observed an absence of explicit criteria for determining whether a framework has been applied successfully. Only a small number of predictive frameworks provided quantitative benchmarks or performance expectations \cite[e.g.,][]{10.1145/3411764.3445563}.

\section{Discussion}
We have explored what HCI researchers mean when they write about frameworks, 
what they contribute when they propose new frameworks, 
how they use frameworks, 
and how they validate frameworks. Next, we discuss the high-level findings from this exploration.

\subsection{The Lifecycle of Frameworks in HCI}
We interpret the six types of engagement identified in our landscape analysis (RQ1) as forming a ``lifecycle'' of frameworks in HCI. In a mature scholarly ecosystem, such a lifecycle would ideally be balanced: New ideas are generated (\Code{Create}), iteratively refined (\Code{Adapt}), rigorously tested (\Code{Validate}), widely applied (\Code{Use} and \Code{Mention}), and ultimately consolidated or historicized (\Code{Review}).

Our findings, however, reveal a lifecycle heavily skewed toward \Code{Create} (54.6\%). In contrast, the practices associated with maturation and consolidation are strikingly rare. The low prevalence of \Code{Adapt} (7.15\%), \Code{Validate} (3.9\%), and \Code{Review} (0.65\%) highlights a substantial gap in iterative refinement, rigorous testing, and systematic reflection on existing frameworks. While this imbalance may be partly attributable to a ``survival bias,'' where novel contributions are favored in the review process, it nonetheless reflects a community practice that prioritizes production over consolidation, posing a challenge for building a cumulative and reliable body of frameworks.

% The community is continuously producing new conceptual tools but invests far less effort in formally iterating or assessing them.
%

Despite this imbalance, our analysis did uncover encouraging signs of a healthy lifecycle in action. We identified a direct lineage in which frameworks first proposed in papers coded as \Code{Create}---namely, the Disclosure Decision-Making framework~\cite{10.1145/3173574.3173732} and the Materials Experience framework~\cite{10.1145/2702123.2702337}---later became the subject of papers coded as \Code{Validate}~\cite{10.1145/3411764.3445331,10.1145/3613904.3642142}. 
Although such cases are rare, they demonstrate that a full lifecycle is possible within the HCI community. These examples serve as positive exemplars of how theoretical contributions can be extended and consolidated over time, pointing toward a path of greater methodological maturity for the field.

% mapping our 6 types to the STS knowledge lifcycle

% frameworks in HCI are not isolated concepts but instead pass through a dynamic cycle of production → consumption → reproduction → reflection → sedimentation.

% What our analysis shows, however, is that framework production is relatively frequent, while validation and reflection are comparatively rare. Likewise, only a small subset of frameworks become widely applied. This uneven distribution may itself reflect the ecological conditions of HCI scholarship. It resonates with Campbell’s evolutionary epistemology and Popper’s three worlds model, both of which emphasize that knowledge is constantly being generated, selectively retained, and preserved.

\subsection{Rethinking the Creation of Frameworks}

Although our analysis shows that \Code{Create} is the dominant mode of engagement, it also raises the question whether creating a framework within HCI guarantees its uptake by the community? Our findings suggest that this is not always the case, pointing to a need to rethink the practice of framework creation itself.

For instance, some HCI-specific frameworks are sometimes bypassed in favor of their original disciplinary sources. Consider intersectionality.  A CHI 2017 paper explicitly positioned intersectionality as a new framework for HCI~\cite{10.1145/3025453.3025766}, drawing from Crenshaw and critical race theory~\cite{crenshaw2013demarginalizing}. 
However, a subsequent CHI 2019 paper on a similar topic~\cite{10.1145/3290605.3300369} applies intersectionality but does not cite the 2017 HCI-specific framework, but rather the foundational sources from social sciences and feminist theory.

Such ``re-importing'' may reflect a preference and typical citation practice for the perceived authority of a foundational theory over a more recent, field-specific adaptation. 
It may also hinder the development of a robust, internal scholarly lineage with in HCI. It makes it difficult to track the evolution of ideas within our community and to build upon each other's conceptual work.

% This leads to a crucial question for our community: 
% What does it take to create a framework that is not just a novel contribution, but one that becomes a foundational and widely-cited artifact within HCI?

% Framework creation does not guarantee uptake inside HCI

% The Intersectional HCI paper (CHI 2017) explicitly positioned intersectionality as a new framework for HCI, drawing from Crenshaw and critical race theory. Yet only two years later, another CHI paper (Plurality of Black Women’s Gameplay Experiences, 2019) applies intersectionality but does not cite the 2017 HCI-specific framework paper. This suggests that even when a framework is introduced within HCI, subsequent work may bypass it and instead cite the original disciplinary sources (here, social sciences and feminist theory).

% Frameworks can be “re-imported” from their parent discipline, bypassing HCI-specific lineage.

% Tracing citations gives insight into whether HCI-created frameworks actually shape subsequent HCI practice, or whether authors prefer to anchor their work in external disciplinary authority.

% are we really creating frameworks ? 

\subsection{Rethinking the Evaluation of Framework}

In this paper, we also examined framework contributions and their normative practices through their functional types, namely \textit{descriptive, explanatory, predictive, prescriptive, and generative}. 
Our finding that generative frameworks often lack formal validation methods connects to a broader debate in HCI. 
The longstanding discussion over whether and how to evaluate conceptual contributions~\cite{Greenberg08usability} may reflect the absence of established methods for appraising more theoretical work~\cite{Salovaara17evaluation}. 
One consequence is that research frequently culminates in the production of conceptual tools such as frameworks. These contributions are undoubtedly valuable and influential within the community, yet they often struggle to find practical application---a phenomenon widely recognized as the ``theory--practice gap''~\cite{Roedl13designresearch}.

Our own analysis offers a potential explanation for why this gap persists and may even be widening. By showing that a majority of frameworks lack explicit application processes and success criteria, we see that the burden of translating a conceptual idea into a practical method is often left entirely to potential adopters, making the leap difficult.

Echoing calls from fields such as Sustainable HCI~\cite{10.1145/3173574.3173790}, we think there is a need for the community to shift its emphasis from purely generating new frameworks to ensuring that they are usable and used (to adapt, validate, and use). 
We suggest that learning from the articulation practices of the most well-defined frameworks can serve as a basis for developing better and more appropriate evaluation methods for all functional types with their corresponding power.

% evaluation methods.

% this paper~\cite{lafon04designing} mentioned about evaluate through descriptive, evaluative, and generative power. 
%10.1145/3544548.3580962 paper guide how to evaluate future EMG system, but they didn't evaluate the step by step framework its self

\subsection{Design Frameworks and Design Implications}

Our review highlights the central role of the \textit{Design Framework} in shaping HCI’s disciplinary identity. Many of these frameworks aim for generative power, seeking to connect research findings to the act of inspiring new designs. This often takes the form of prescriptive design recommendations or open-ended implications.

However, since there is no well-established tool for evaluating generative power, this practice emphasis on producing direct design implications may be a point of tension within the field. 
As Dourish has argued~\cite{Dourish06implications}, when ethnography research is forced to generate a simple list of design recommendations, the true value of insight can be lost and the recommendations themselves may devalue the creative and analytical work central to the design process. 
We share this view. A design framework does not have to focus solely on design implications, and its generative power is not its only functional goal.
The design framework's primary goal can also be to equip designers with a richer, more nuanced perspective on a problem space with the framework's descriptive, explanatory, predictive, or prescriptive power. 
 We see encouraging signs of this in our corpus, as we found that most papers proposing a Design Framework also include key components that are descriptive or explanatory. And we see more conceptual framework in our field. 
 This suggests that the community is already moving towards creating more holistic design tools that enrich the designer’s own creative and analytical process.

\subsection{\rev{What Really Counts as a Framework}}
\rev{
At this stage, the reader might wonder what a framework really is, particularly given the promise of the title and the understanding that the meaning of a framework depends on how authors employ the term (see \autoref{sec:authorcentric}). From the analysis, several functions of frameworks have become clearer. A primary characteristic is that frameworks \textit{structure}: they organize phenomena temporally, conceptually, or logically; break tasks into steps; or delineate facets of analysis. This view is more specific than the definition offered by \citet{roger12HCItheory} that was quoted in the introduction. Furthermore, this structuring function helps to unify the different types presented in \autoref{tab:labeltype}.}

\rev{Despite this common emphasis on structure, frameworks are typically characterized by having a broad scope that affords substantial \textit{interpretive freedom}. This distinguishes their application from that of theories or constructs, which are often more narrowly defined and operationalized. Good frameworks, particularly those imported from outside HCI, support multiple interpretations while still making explicit their fundamental assumptions and commitments.}

\rev{Finally, it is important to recognize that frameworks are intended to \textit{guide researchers}. In our view, this function is not sufficiently appreciated. Ideally, frameworks should prescribe clear processes for their application, provide examples of their use, and highlight common pitfalls. Yet in many cases, this aspect of framework design remains underdeveloped or overlooked. Thus, the many frameworks that are proposed in CHI often do not include enough guidance, neither in their original formulation or in later adaptations or uses.}

\subsection{\rev{Guidance for Proposing and Using Frameworks}}
\rev{Based on the preceding analysis and discussion, we wish to move beyond describing how frameworks are proposed and used, and instead offer guidance on how researchers might do so more effectively. Such guidance is necessarily preliminary, yet may nevertheless be valuable.}

\revtwo{
When \textit{proposing frameworks}, we offer five initial recommendations:
\begin{itemize}
    \item Articulating the benefits of the framework being proposed more explicitly---particularly when it is imported from outside HCI---and explain why existing frameworks cannot be adapted to achieve those benefits. The reason for this recommendation is that we seems to be much better at proposing frameworks than adapting or just using existing ones.
    \item Identifying primary functional roles. Rather than using ``framework'' as a vague label, authors should explicitly identify their work's primary roles --- selecting one or two dominant types from our functional typology (e.g., Descriptive for organizing phenomena or Explanatory for causal relationships). Narrowing this focus reduces the ambiguity currently found in the community’s lexicon. 
    \item Focusing on core constructs that align with the framework’s functional role. For example, a descriptive framework may adopt a shared vocabulary as its core constructs to characterize an emerging interaction context. Authors should iteratively refine these constructs to ensure they offer a clear, generalizable, and extensible structure. To reduce readers’ cognitive effort, it is important to select visualization techniques that match the underlying construct logic—such as matrices for organizing dimensions and flow diagrams for representing causal or step-by-step processes. In our corpus, many proposed frameworks lack readily identifiable core constructs, which makes them difficult to analyze; correspondingly, readers may also struggle to efficiently extract key insights.
    \item Incorporating a validation roadmap. This does not imply that every framework must be validated in its initial version, nor should the need for validation suppress novel ideas. 
    However, providing at least preliminary evidence that a framework works in its intended functional role is essential. In addition, authors should explicitly articulate the framework’s underlying assumptions and outline plausible scenarios or contexts in which its utility can be examined or tested. In our corpus, very few frameworks report any form of validation or even a concrete plan for validation.
    This stands in notable contrast to how rarely we accept unvalidated claims about systems or empirical findings.
    \item Providing explicit instructions for reuse. A central function of an HCI framework is to guide other researchers toward action. However, our findings suggest a prevalent ``articulation gap'': while authors often demonstrate their framework by applying it to their own data, they rarely provide explicit instructions for how others should use it in different contexts. 
    The lack of such guidance may partly explain why so many new frameworks are created: it is often unclear how existing ones should be used. To lower the barrier for community adoption, authors should move beyond implicit demonstration and clearly articulate concrete instructions, considerations, or checklist for applying their frameworks in new settings. 
\end{itemize}
}

% \rev{When \textit{proposing frameworks}, we offer three initial recommendations.
% First, researchers should more explicitly articulate the benefits of the framework being proposed---particularly when it is imported from outside HCI---and explain why existing frameworks cannot be adapted to achieve those benefits. The reason for this recommendation is that we seems to be much better at proposing frameworks than adapting or just using exisiting ones.
% Second, validation of frameworks is crucial. This does not imply that every framework must be validated in its initial version, nor should the need for validation suppress novel ideas. However, in our sample, fewer than 5\% of frameworks reported any form of validation. This stands in notable contrast to how rarely we accept unvalidated claims about systems or empirical findings.
% Third, because a central function of frameworks is to guide researchers to action, authors should provide clearer instructions for how users are meant to apply the framework. The lack of such guidance may partly explain why so many new frameworks are created: it is often unclear how existing ones should be used.}

\rev{When \textit{using frameworks}, a similar set of preliminary guiding questions can be identified.
First, the manner in which a framework is applied is crucial, not only for its usefulness in a given study but also for what the community can learn from its application. This should be reported much more frequently and in more detail.
Second, when frameworks are imported from outside HCI, researchers should be explicit about the benefits they bring and about the nature of the translation involved in adapting them for HCI contexts. In particular, be clear about underlying assumptions and commitments that bound application, despite the work required in interpreting the framework.
Third, researchers should aim to related their use of the framework to its core ideas, recommendations, and guidance for use. Much as scholars write about ``talking back to theory'' \cite{beck2016examining}, it would be valuable for the community to more routinely discuss and reflect on experiences with applying frameworks. This would help other researchers but also allow for better reviews of frameworks and perhaps also provide ideas for their adaptation.}

\section{Limitations}

We also acknowledge the boundaries and limitations of this work. 
First, our analysis is necessarily limited to an author-centric view as presented in the text. 
We do not capture how a framework's users (other researchers and practitioners) or its evaluators (paper reviewers) perceive its function and power. 
A framework's intended function may not fully encompass its eventual use by the community. 
For example, while the Trajectories Framework~\cite{Benford08trajectories} was proposed primarily as a conceptual tool, a later citation analysis by Velt et al. ~\cite{veltSurveyTrajectoriesConceptual2017} revealed that it was used in ways that reflected all five functional types.
This suggests that a framework's contribution is co-constructed by its adopters and highlights an opportunity for future work to understand this reception and reinterpretation process.

Second, we chose CHI as a single, representative venue for HCI. However, other conferences or journals (such as CSCW, UIST, or DIS) may have different conventions regarding the creation, articulation, and validation of frameworks, meaning the practice norms we observed may be nuanced. 
Furthermore, by analyzing a large and diverse corpus spanning a decade, our study prioritizes breadth over depth. 
While this approach is effective for identifying high-level patterns, it may not capture the nuances of individual contributions. 
We envision future work that takes a more focused perspective, perhaps by deeply analyzing one certain type of framework or the practices within a specific sub-community.

\rev{Third, our reliance on what authors consider a framework does not preclude false negatives: Papers where authors suggest something that functions as a framework but is called something else. Our rationale for this decision was given in \autoref{sec:authorcentric}. Future work should explore the extent and possible content of false negatives.}

\rev{Fourth, our characterisation of framework use is based on text analysis. An alternative method with different strengths and weaknesses is citation analysis. We would like to see future work provide additional evidence for our claims by analysing how framework papers are cited.}

\section{Conclusion}

In this paper, we review 10 years of CHI papers (2015-2024, N = 615) that use the term framework to understand its role and associated practices.
Our investigation reveals a vibrant and rapidly growing body of framework-related scholarship that is simultaneously productive and imbalanced. 
We classify the papers into six types of engagement that form a lifecycle of framework contribution in HCI. 
We find a landscape dominated by the creation of new frameworks with far less emphasis on adapting, validating, or reviewing existing ones. 
Although using frameworks occurs frequently, we also found that newly proposed frameworks are rarely adopted and used within our corpus, suggesting a gap between production and uptake.
Focusing on frameworks presented as one of the contributions, we ground their characteristics in a functional typology, exploring the descriptive, explanatory, predictive, prescriptive, and generative roles they play. 
Finally, our analysis of methodological practices highlights two critical gaps: A lack of systematic validation for new contributions and a lack of clear articulation to guide their reuse.

\begin{acks}
This work was supported in part by the Villum Fonden research grant (VIL73783, TRACTION). Additional support was provided by JST ASPIRE for Top Scientists (Grant Number JPMJAP2405) and JST SPRING (Grant Number JPMJSP2108).
\end{acks}

\bibliographystyle{ACM-Reference-Format}
\bibliography{reference}

\appendix
\section{Decision Protocol for Classification}
\label{app:decision_tree}

To ensure each paper was classified consistently, we employed the following structured decision tree. For each paper, the coder answered a sequence of questions to determine the single most appropriate code, prioritizing more substantive engagements. 

\begin{enumerate}
    \item \textbf{Q1}: Does the paper explicitly claim to propose a new framework (e.g., ``we propose a framework...'', ``our framework'')?
    \begin{itemize}
        \item If YES, proceed to Q2.
        \item If NO, proceed to Q3.
    \end{itemize}
    
    \item \textbf{Q2}: Is this claimed framework clearly derived, extended, or tailored from an existing, named ``framework'' (not a theory or model)?
    \begin{itemize}
        \item If YES, the paper is coded as \textbf{\Code{Adapt}}.
        \item If NO, the paper is coded as \textbf{\Code{Create}}.
    \end{itemize}
    
    \item \textbf{Q3}: Does the paper's primary activity involve analyzing, testing, evaluating, or conceptually critiquing a specific, existing framework?
    \begin{itemize}
        \item If YES, proceed to Q4.
        \item If NO, proceed to Q5.
    \end{itemize}
    
    \item \textbf{Q4}: Is the paper analyzing how *others* have used the framework (e.g., via a literature review or meta-analysis)?
    \begin{itemize}
        \item If YES, the paper is coded as \textbf{\Code{Review}}.
        \item If NO, the paper is coded as \textbf{\Code{Validate}}.
    \end{itemize}
    
    \item \textbf{Q5}: Does the paper apply an existing framework as a primary tool or lens to guide its design, analysis, or discussion?
    \begin{itemize}
        \item If YES, the paper is coded as \textbf{\Code{Use}}.
        \item If NO, the paper is coded as \textbf{\Code{Mention}}.
    \end{itemize}
\end{enumerate}

\section{Detailed Coding Principles}
\label{app:coding_principles}

This appendix details the specific rules derived from our core principle of \textit{terminological literalism} and our \textit{author-centric} view. These rules guided our classification process.

\begin{enumerate}
    \item A contribution was coded as \textbf{\Code{Create}} only if the authors explicitly identified their primary contribution as a ``framework.'' 
    If they used other term to summarize their contribution, like ``model'' or ``theory'', the \textbf{Q1} will be NO. 
    
    % Other forms of conceptual contributions, such as models or theories, were not classified as such, regardless of their functional similarity.
    
    \item This rule extended to the \textbf{\Code{Adapt}} code. A paper was classified as \Code{Adapt} only if the authors explicitly stated their contribution was derived from or an extension of another specific, pre-existing ``framework.'' 
    If a contribution was described as being built upon a ``theory'' or ``model,'', the Q2 would go to No and it was coded as \Code{Create} rather than \Code{Adapt}.
    
    \item Conversely, for the \textbf{\Code{Use}} code, we respected the applying author's framing. 
    If a paper claimed to use an existing piece of knowledge ``as a framework'' for its analysis or design process, we coded it as \Code{Use}, even if the original cited work was not itself presented as a framework.
\end{enumerate}

\section{Pipeline for Clustering Themes}
\label{app:clustering}
\rev{
We employed a computational pipeline combined with qualitative refinement to identify latent thematic structures. First, we generated contextual embedding vectors for each paper's keywords using the SentenceTransformer~\cite{reimers2019sentence} model (all-MiniLM-L6-v2\footnote{\url{https://huggingface.co/sentence-transformers/all-MiniLM-L6-v2}}). 
We then performed preliminary clustering using \textit{K-Means}, initializing the number of clusters via a heuristic ($k \approx N/50$) with a fixed random seed (42) to ensure reproducibility. To facilitate visual assessment and interpretability of the high-dimensional embedding space, we utilized UMAP (Uniform Manifold Approximation and Projection)~\cite{mcinnes2018umap} to project the data into two dimensions ($n\_neighbors=30$, $min\_dist=0.05$, metric=\textit{cosine}).
} 

\rev{
To ensure the meaningfulness of these computational clusters, we applied an iterative human-in-the-loop refinement process.
We developed custom scripts to extract and summarize high-frequency keywords for each potential cluster. 
Based on these summaries, we conducted multiple rounds of hyperparameter tuning and manually inspected samples with lower semantic similarity to resolve ambiguities. 
We iteratively merged semantically overlapping groups and refined cluster boundaries to ensure distinctness.}
\rev{
This process resulted in 9 distinct thematic clusters (Figure~\ref{fig:domaincluster}) that describe the primary domains of framework engagement. 
}

\section{Complete Paper List}

\revtwo{The full list of papers and coding results is available in the supplementary materials and the accompanying GitHub repository (\url{https://github.com/First2nd3rd/frameworks-in-hci}).
}

\end{document}